\documentclass[reprint,superscriptaddress,
 amsmath,amssymb,
pra,
floatfix,
]{revtex4-1}
\usepackage[utf8]{inputenc}

\usepackage{graphicx}% Include figure files
\usepackage{dcolumn}% Align table columns on decimal point
\usepackage{bm}% bold math
\usepackage{hyperref}% add hypertext capabilities
\usepackage{ulem}
\usepackage{xcolor}
\usepackage{amsmath, amssymb}
\usepackage[T1]{fontenc}

\usepackage{braket}
\renewcommand{\vec}[1]{{\boldsymbol{#1}}} % for vectors

\begin{document}
	
	\preprint{APS/123-QED}

	%\preprint{This line only printed with preprint option}
	
	\title{Dynamics of equilibration and collisions in ultradilute quantum droplets}

	\author{V. Cikojevi\'c}
	\affiliation{University of Split, Faculty of Science, Ru\dj era Bo\v{s}kovi\'ca 33, HR-21000 Split, Croatia}
	\affiliation{Departament de F\'{\i}sica, Universitat Polit\`ecnica de Catalunya, Campus Nord B4-B5, E-08034 Barcelona, Spain}
	
	\author{L. Vranje\v{s} Marki\'c}
	\affiliation{University of Split, Faculty of Science, Ru\dj era Bo\v{s}kovi\'ca 33, HR-21000 Split, Croatia}
	
	\author{M. Pi}
	\affiliation{Departament FQA, Facultat de F\'{\i}sica, Universitat de Barcelona, Diagonal 645, 08028 Barcelona, Spain}
	\affiliation{Institute of Nanoscience and Nanotechnology (IN2UB), Universitat de Barcelona, 08028 Barcelona, Spain}
	
	\author{M. Barranco} 
	\affiliation{Departament FQA, Facultat de F\'{\i}sica, Universitat de Barcelona, Diagonal 645, 08028 Barcelona, Spain}
	\affiliation{Institute of Nanoscience and Nanotechnology (IN2UB), Universitat de Barcelona, 08028 Barcelona, Spain}

	\author{F. Ancilotto}
	\affiliation{Dipartimento di Fisica e Astronomia "Galileo Galilei" and CNISM, Universita di Padova, via Marzolo 8, 35122 Padova, Italy}
	\affiliation{CNR-IOM Democritos, via Bonomea 265, I-34136 Trieste, Italy}

	\author{J. Boronat} 
	\affiliation{Departament de F\'{\i}sica, Universitat Polit\`ecnica de Catalunya, Campus Nord B4-B5, E-08034 Barcelona, Spain}

	\date{\today}% It is always \today, today,
	
	\begin{abstract}
		Employing time-dependent density-functional theory, we have studied 
dynamical equilibration and binary head-on collisions of quantum droplets made of a 
$^{39}$K-$^{39}$K Bose mixture. The phase space of collision outcomes is 
extensively explored by performing fully three-dimensional calculations with 
effective single-component QMC based  and two-components LHY-corrected mean-field functionals. We exhaustively explored the important effect --not considered in previous studies-- of the initial population ratio deviating from 
the optimal mean-field value $N_2/N_1 = \sqrt{a_{11} / a_{22}}$. Both 
stationary and dynamical calculations with an initial non-optimal concentration 
ratio display good agreement with experiments.  Calculations including 
three-body losses acting only on the $\ket{F, m_F} = \ket{1,0}$ state show 
dramatic differences  with those obtained with the  three-body term 
acting on the total density. 
	\end{abstract}
	
	%\pacs{}% PACS, the Physics and Astronomy

	%\date{\today}
	\maketitle
	%{\let\newpage\relax\maketitle}

	\section{Introduction}
	
%\mbgcomm{WHEN CITED TOGETHER, REFS. SHOULD BE TIME-ORDERED; SEE E.G. ORIGINAL REFS 4-7}\\

The collision of liquid drops  is one of the more fundamental and complex problems  addressed in fluid dynamics, with implications in basic 
research and applications  e.g. in microfluidics, formation of rain drops,  ink-jet printing, or spraying for combustion, painting and 
coating \cite{ashgriz1990coalescence,qian1997regimes,Pan09,Var16}. 
Liquid drop collisions were also used as a model for nucleus-nucleus reactions \cite{Swi82} and nanoscopic 
$^3$He droplets collisions \cite{Gui95}.

Generally speaking, upon collision droplets may bounce back, coalesce into a single drop, separate after temporarily forming a 
partially fused system,  
or shatter into a cloud of small droplets. The main goal of the studies on droplet-droplet collisions is to determine how the appearance of these 
regimes depends on the collision parameters  (droplet size, velocity, impact parameter) and intrinsic properties of the 
liquid (viscosity, surface tension, ...).

With the advent of the so-called ``helium drop machines'', it has been possible to generate
$^4$He nanodroplets by the free expansion of a supercooled gas, as reviewed e.g. 
in Ref.~\cite{toennies2004superfluid}.
This has allowed to extend the  study of  liquid droplets to the quantum regime. 
Indeed, helium in its two isotopic forms, $^3$He (a fermion) and $^4$He (a boson),
is the only element in nature which may exist at zero temperature as an extended 
liquid or in the form of droplets.  
Superfluid $^4$He samples are ideal systems to study quantized vortices  --a 
most striking hallmark of superfluidity \cite{Don91}--  and quantum turbulence \cite{Tsu09}.

At the experimental droplet temperatures, 0.37 K for $^4$He and 0.15 K for $^3$He\cite{toennies2004superfluid}, $^4$He is  a 
superfluid and $^3$He is a normal fluid. Studies on superfluid $^4$He droplets collisions are very scarce, see e.g. 
Refs.~\cite{vicente2000coalescence,maris2003analysis,ishiguro2004coalescence,
escartin2019vorticity}. 
 It has been recently shown that the merging of 
superfluid $^4$He nanodroplets may produce quantized vortices and surface turbulence \cite{naz15,escartin2019vorticity}

In the field of ultracold Bose gases, there are also very few studies of cold
gas collisions \cite{sun2008slow, 
astrakharchik2018dynamics,zhong2021oscillatory}, since the only accessible phase 
until recently was the gaseous, confined one. Nevertheless, there are many 
interesting dynamical phenomena \cite{pitaevskii2016bose} whose description has 
attracted interest, such as the study of collective 
excitations in confined Bose and Fermi systems, with different dimensionalities, 
in gaseous \cite{dalfovo1999theory,jin1996collective,Pollack2010collective, 
skov2020lhy,Atas2017,rosi2015collective,Kinast2004evidence, 
Bartenstein2004collective, altmeyer2007precision}, liquid 
\cite{tylutki2020collective, hu2020collective} or supersolid regimes 
\cite{tanzi2019supersolid, leonard2017monitoring}.

The situation has changed quite recently since 
a self-bound, low-density liquid-like  state composed by ultracold quantum Bose-Bose mixtures, first 
theoretically proposed by Petrov \cite{petrov2015quantum}, has been experimentally produced 
\cite{cabrera2018quantum,cheiney2018bright,semeghini2018self, 
derrico2019observation}. In these mixtures, an adequately tuned 
interaction can lead to a regime where the mean-field energy is comparable to 
the Lee, Huang, and Yang (LHY) energy.  The LHY energy term is a 
perturbative correction to the mean-field energy, first calculated with the single-component Bose gas \cite{lee1957many,huang1957quantum}, and later 
extended to two-component Bose-Bose mixtures 
\cite{larsen1963binary,minardi2019effective,naidon2021bubbles}. This desirable feature of 
stabilizing the mean-field collapse in an LHY-extended theory (MF+LHY), allowing 
for droplet formation, is present also in low-dimensional geometries 
\cite{petrov2016ultradilute,parisi2019liquid, parisi2020quantum} and accounts 
for the stability of dipolar droplets \cite{bottcher2019dilute,bottcher2021}. Consequently, 
the realm of stable quantum droplets has been extended to densities much lower 
than those of helium droplets \cite{barranco2006helium}.
	
	As pointed out in Ref.~\cite{hu2020consistent}, the LHY term suffers from an
	intrinsic inconsistency with the appearance of an imaginary term in 
	the 
energy of the Bose-Bose mixture. This is at variance with a single-component Bose gas, 
where 
no such imaginary term appears, and where the  LHY term 
has proven to be valid up to relatively large densities, as it was confirmed by  
first-principle Quantum Monte Carlo (QMC) calculations \cite{giorgini1999ground, 
rossi2013path}. Already in the first experimental realization of a quantum 
Bose-Bose liquid mixture, composed by two hyperfine states of $^{39}$K, 
there have been deviations of observed properties 
from the predictions of usually employed theory 
based on interactions described solely in terms of $s$-wave scattering 
lengths
\cite{petrov2015quantum}. These effects  were properly explained instead 
by a QMC-based 
functional built to include effective-range corrections 
\cite{cikojevic2020finite}. 
Diffusion Monte Carlo (DMC) calculations indicate that inclusion of
the effective range allows to extend the universality of the theory 
\cite{cikojevic2019universality,cikojevic2020finite}, providing 
improved energy functionals when both the $s$-wave 
scattering lengths and the effective ranges are known.
		
	Some progress in the understanding of dynamical properties of quantum 
Bose-Bose droplets has been made in a recent experimental study of head-on collisions 
between pairs of $^{39}$K-$^{39}$K droplets
\cite{ferioli2019collisions}, providing a new avenue of research. In Ref. 
\cite{ferioli2019collisions}, the authors discovered a highly compressible 
droplet regime, not present in the world of classical liquids, when the total 
atom numbers in colliding droplets are small. Thus, it is not clear whether the Weber 
number theory \cite{frohn2000dynamics}, describing the dynamics of classical 
liquid collisions,  can apply to ultra-dilute droplets. Depending on the 
velocity of each droplet, the outcome of the experiment 
\cite{ferioli2019collisions} was either merging or separation of
the colliding droplets.

As in Ref. \cite{ferioli2019collisions}, we define a critical velocity $v_c$ as 
the initial velocity of each droplet above (below) which  the two droplets 
separate (merge) upon colliding. 
A significant discrepancy  was observed between the experimental 
critical velocity $v_c$ and the theoretical analysis of the
experiment carried out within the MF+LHY 
approach \cite{ferioli2019collisions}. The disagreement was attributed to the lacking of  three-body losses (3BL), which
were introduced in a next step but acting on the total density. This means that both components lose their atoms such that 
the density ratio is constantly kept fixed at the value $\rho_2 / \rho_1 = \sqrt{a_{11} / 
a_{22}}$, which is the mean-field stability requirement \cite{petrov2015quantum,staudinger2018self,ancilotto2018self}. 
By doing so, the agreement between theory and experiment was found to be good.
However, it is unclear whether this procedure 
masks other interesting details which might be hidden by the fact that 3BL are made to act on the total density and not 
on the appropriate component of the bosonic mixture. 
In particular, the procedure excludes the possibility that the droplets in the experiment are not  fully equilibrated. 
Additionally, experimental measures \cite{semeghini2018self} have demonstrated significant differences in intensities of 3BL of different components, which are not taken into account by the approach of Ref. \cite{ferioli2019collisions}.
In this work, we re-analyze that experiment  by using a 
QMC-based functional which takes into account the effective range of the interactions and a two-component MF+LHY functional that enables the study of unequilibrated drops and an experimentally more realistic consideration of 3BL. Our results show 
the relevance of the non-optimal concentration ratio in the outcome of drop 
collisions, providing an alternative explanation to the experimental results 
without invoking 3BL. 	
	
This work is organized as follows. In Sec. \ref{sec:mflhy_eos} we lay 
out the basic equations of the extended LHY mean-field theory (MF+LHY). In 
Sec. \ref{sec:simulation_details} we present the details of our simulations. 
In Sec. \ref{sec:stationary_drop_calc} we discuss the effects of 
the non-optimal initial atom number ratio and 3BL on the stationary 
drop. In Sec. \ref{sec:eff_single_comp_results}, we systematically compare the 
drops collisions results obtained within the effective single-component  MF+LHY theory 
with those obtained using the QMC-based functional. In 
Sec. \ref{sec:two_comp_calculations} we report results derived with the 
two-component framework. In Sec. \ref{sec:three_body_losses} we investigate the influence on the collisions of both the initial population imbalance and 3BL acting only on the $\ket{F, m_F} = 
\ket{1,0}$ state. Finally,  Sec. \ref{sec:conclusions} 
comprises the main conclusions of our work.

	\section{\label{sec:mflhy_eos}MF+LHY and QMC density functionals}

	The LHY-extended mean-field theory (MF+LHY) is 
based on the following density functional per unit volume $V$
\cite{petrov2015quantum,ancilotto2018self},
	\begin{equation}
	\label{eq:two_comp_lhy}
	\mathcal{E}[\rho_1, \rho_2] = E_{\rm MF}/V + E_{\rm LHY} / V \ ,
	\end{equation}
	where
	\begin{equation}
	\dfrac{E_{\rm MF}}{V} =   \dfrac{2\pi\hbar^2 a_{11}}{m} \rho_1^2 + 
\dfrac{2\pi\hbar^2 a_{22}}{m} \rho_2^2 +  \dfrac{4\pi\hbar^2 a_{12}}{m}\rho_1 
\rho_2 \ ,
	\end{equation}
	and
	\begin{equation}
	\dfrac{E_{\rm LHY}}{V}  =  \dfrac{256 \sqrt{\pi} \hbar^2}{15 m} (a_{11} \rho_1)^{5 / 2} f\left(1, 
\dfrac{a_{12}^2}{a_{11} a_{22}}, \dfrac{a_{22} \rho_2}{a_{11} \rho_1}\right) \ .
	\end{equation}
	We have considered equal masses $m_1 = m_2 = m$ and the function
	$f$ is defined in Ref. \cite{petrov2015quantum},

	\begin{eqnarray}
	\label{eq:definition_f}
	f(1, u,x) =\dfrac{1}{{4\sqrt{2}}}[(1 + x + \sqrt{(1-x)^2 + 4ux})^{5/2} \nonumber\\
	+ (1 +x - \sqrt{(1-x)^2 + 4ux})^{5/2}] \ .
	\end{eqnarray}
Here $u={a_{12}^2}/{(a_{11} a_{22})}$ and $x={a_{22} \rho_2}/{(a_{11} \rho_1)}$.
	The function $f$ is complex for $a_{12}^2 > a_{11} a_{22}$, and we use the 
approximation ${a_{12}^2} = {a_{11} a_{22}}$ to keep $f$ real. By doing so, 
$f(1, 1, x) = (1+x)^{5/2}$ and the LHY functional reads
	\begin{equation}
	\dfrac{    E_{\rm LHY}}{V}  =  
\dfrac{256\sqrt{\pi} \hbar^2}{15 m} \left(a_{11} \rho_1 + 
a_{22} \rho_2\right)^{5/2}.
	\end{equation}        
	The above framework is the most general version of a two-component 
equal-masses Bose-Bose energy functional, and it 
allows for all possible $\rho_1$ and $\rho_2$ values. This functional can be 
reduced to an effective one-component functional, which is the one mostly used 
in the study of Bose-Bose mixtures, if one 
uses the result that the stability of a dilute Bose-Bose 
 mixture lies in a very narrow range of optimal partial densities $\rho_1 / 
\rho_2 = \sqrt{a_{22}/a_{11}}$ \cite{staudinger2018self, ancilotto2018self}. 
Then, for this fixed ratio of $\rho_1/ \rho_2$ the MF+LHY theory can be written in the compact form 
	\begin{equation}
	\label{eq:mflhy_eos}
	\dfrac{E/N}{|E_0|/N} = -3\left(\dfrac{\rho}{\rho_0}\right)+ 2 
\left(\dfrac{\rho}{\rho_0}\right)^{3/2} \ ,
	\end{equation}
	with $\rho$ being the total density and $E_0 / N$ and $\rho_0$ the equilibrium 
energy per particle and equilibrium total density, respectively, given by
	\begin{equation}
	\label{eq:rho_0}
	\rho_0 = \dfrac{25 \pi}{1024} \dfrac{\left(a_{12}/a_{11} + 
\sqrt{a_{22}/a_{11}}\right)^2}{\left(a_{22}/a_{11}\right)^{3/2}\left(1+\sqrt{a_{
22}/a_{11}}\right)^4}   \dfrac{1}{a_{11}^3} \ ,
	\end{equation}    
	\begin{equation}
	\dfrac{E_0}{N} = \dfrac{-\hbar^2}{ma_{11}^2}\dfrac{25\pi^2}{768} 
\dfrac{|a_{12}/a_{11} + \sqrt{a_{22}/a_{11}}|^3}{a_{22}/a_{11} (1+ 
\sqrt{a_{22}/a_{11}})^6} = \dfrac{-\hbar^2}{2m \xi^2} \ .
	\end{equation}
	We have introduced in the last equation the healing length $\xi$, 
obtained by equating the kinetic energy to the energy per 
particle at the equilibrium density: 
	\begin{equation}
	\label{eq:healing_length}
	\dfrac{\xi}{a_{11}} =  \dfrac{8\sqrt{6}}{5\pi} \sqrt{\dfrac{a_{22}}{a_{11}}} 
\dfrac{(1 + \sqrt{a_{22} / a_{11}})^3}{|a_{12}/a_{11} + \sqrt{a_{22} / 
a_{11}}|^{3/2}} \ .
	\end{equation}
	The total atom number in 
	%, with respect to 
	the reduced unit $\tilde{N}$ 
introduced in Ref.~\cite{petrov2015quantum}, is given by
	\begin{equation}
	\label{eq:atom_number_mflhy}
	\dfrac{N}{\tilde{N}} = \dfrac{3\sqrt{6}}{5\pi^2} 
\dfrac{\left(1+\sqrt{a_{22}/a_{11}}\right)^5}{|a_{12}/a_{11} + 
\sqrt{a_{22}/a_{11}}|^{5/2}} \ .
	\end{equation}
%	Finally, t
	To compare with the results of Ref.~\cite{ferioli2019collisions}, 
the velocity $v$ can be expressed in the universal unit $\tilde{v}$ as
	\begin{equation}
	\label{eq:velocity_mflhy}
	\dfrac{v}{\tilde{v}} = \dfrac{5\pi \hbar}{8\sqrt{6} m a_{11}} \dfrac{ 
|a_{12} / a_{11} + \sqrt{a_{22} / a_{11}}|^{3/2}}{\sqrt{a_{22} / a_{11}} (1 + 
\sqrt{a_{22} / a_{11}})^3} \ .
	\end{equation}

We have also used a density functional derived from QMC 
calculations
(QMC functional in the following), which is 
constructed by performing DMC calculations
of a $^{39}$K mixture in the homogeneous phase 
\cite{cikojevic2020finite}. The QMC functional $\mathcal{E}_{\rm int}$ is obtained with the relation
\begin{equation}
	\mathcal{E}_{\rm int} = \rho \dfrac{E}{N},
\end{equation}
 where $E/N$ is the energy per particle of the extended system, calculated from 
QMC. QMC calculations were performed with the model potentials which reproduce 
both the $s$-wave scattering lengths \textit{and} effective ranges, which are 
known from the experiment \cite{tanzi2018feshbach}. In this way, the QMC 
functional correctly incorporates the two relevant scattering parameters of this 
mixture, i.e., the $s$-wave scattering lengths and the effective ranges.

\medskip

	\section{\label{sec:simulation_details}Time-evolution equations}
	
	For a two-component system, our ansatz for the many-body wave function is
	\begin{equation}
	\label{eq:two_component_ansatz}
	\Psi(\mathbf{r}_1,  \ldots, \mathbf{r}_N;t) = \prod_{i=1}^{N_1} \psi_1(\mathbf{r}_i, t)\prod_{j=1}^{N_2} \psi_2(\mathbf{r}_j, t),
	\end{equation}
	where the number of particles of component 1 (2) is equal to $N_1$ ($N_2$).  
The equations of motion for the two components read
	\begin{equation}
	\label{eq:twocomp_equation_of_motion}
	i\hbar \dfrac{\partial \psi_i}{\partial t}  = \mathcal{H}_i \psi_i = \left\{ 
-\dfrac{\hbar^2}{2m} \nabla^2 + V_i(\rho_1, \rho_2) \right\} \psi_i \ ,
	\end{equation}
	for $i=1,2$, where $\rho_i=|\psi_i|^2$, and the potential $V_i$ is obtained from Eq. (\ref{eq:two_comp_lhy}) \cite{barranco2017zero}
	\begin{equation}
	V_i = \dfrac{\partial \mathcal{E}[\rho_1, \rho_2]}{\partial \rho_i} \ .
	\end{equation}
	Explicitly, the coupled equations of motion for  the two components of the condensate read
	\begin{eqnarray}
	i \hbar \frac{\partial \psi_{1}}{\partial t}
	& =  &
	\left(
	-\frac{\hbar^2}{2 m} \nabla^2
	+
	\dfrac{4\pi\hbar^2 a_{11}}{m} \rho_1  \right.    +   \dfrac{4\pi\hbar^2 a_{12}}{m} \rho_2 + \nonumber \\ & & %\hspace{-1.5cm}
	\frac{128 \sqrt{\pi} \hbar^{2} a_{11}}{3 m}\left(a_{11} \rho_{1}+a_{22} \rho_{2}\right)^{3 / 2}
	\bigg) 
	\psi_{1}   \ ,
	\label{gp1}
	\end{eqnarray}
	\begin{eqnarray}
	i \hbar \frac{\partial \psi_{2}}{\partial t}
	& = & 
	\left(
	-\frac{\hbar^2}{2m} \nabla^2
	+
	\dfrac{4\pi \hbar^2 a_{22}}{m} \rho_2 \right.
	+ \dfrac{4\pi \hbar^2 a_{12}}{m} \rho_1 + \nonumber \\
	&  & \frac{128 \sqrt{\pi} \hbar^{2} a_{22}}{3 m}\left(a_{11} \rho_{1}+a_{22} \rho_{2}\right)^{3 / 2}
	\bigg) 
	\psi_{2} \ .
	\label{gp2}
	\end{eqnarray}    
	When the ratio of densities is the optimal one, $\rho_2 / \rho_1 = 
\sqrt{a_{11} / a_{22}}$, fixed by the condition of minimum energy per particle 
\cite{petrov2015quantum,staudinger2018self,ancilotto2018self}, the energy 
functional for the total density $\rho=\rho_1+\rho_2$ reduces to
	\begin{equation}
	\label{eq:single_comp_functional}
	\mathcal{E}_{\rm int} = \alpha\rho^2 + \beta\rho^{\gamma + 1},
	\end{equation}
	where $\alpha$, $\beta$ and $\gamma$ are either the MF+LHY parameters from 
Eq. (\ref{eq:mflhy_eos}), or the ones which better fit the  DMC equation of 
state \cite{cikojevic2020finite}. In this case, the problem is thus effectively 
single-component, meaning that the full many-body wavefunction 
(\ref{eq:two_component_ansatz}) reduces to
	\begin{equation}
	\Psi(\mathbf{r}_1, \vec{r}_2, \ldots, \mathbf{r}_N; t) = \prod_{i=1}^{N} \psi(\mathbf{r}_i; t),
	\end{equation}
	where $\psi$ is the solution of the following equation
	\begin{equation}
	\label{eq:eq_of_motion_singlecomp}
	i\hbar\dfrac{\partial \psi}{\partial t} = \mathcal{H}\psi 
=\left(\dfrac{-\hbar^2}{2m}\nabla^2 + V(\rho)\right)\psi \ ,
	\end{equation}
	where
	\begin{equation}
	V(\rho) = 2\alpha \rho +  \beta (\gamma + 1)\rho^\gamma \ .
	\end{equation}
	Equations  (\ref{gp1}), (\ref{gp2}) and 
(\ref{eq:eq_of_motion_singlecomp}) are solved numerically by successively 
applying the time-evolution operator
	\begin{equation}
	\label{eq:time_evolution}
	\psi(t + \Delta t) = e^{-i \mathcal{H} \Delta t} \psi(t) \ ,
	\end{equation}
	where a Trotter decomposition accurate to second order in the time step is  
implemented \cite{chin2009any}
	\begin{equation}
	e^{-i \mathcal{H} \Delta t / \hbar}  = e^{-i \Delta t V(\mathbf{R}') / 
2\hbar}  e^{-i\Delta t \hat{K} / \hbar } e^{-i\hbar \Delta t V(\mathbf{R})/ 
2\hbar} + \mathcal{O}(\Delta t^3) \ ,
	\end{equation}
	with $\hat{K} = -\hbar^2 \nabla^2 / (2m)$. The kinetic energy propagator is 
evaluated in $k$-space by means of Fourier transforms.

Since an important parameter in our study is the deviation with respect to the optimal atom ratio $N _2/N_1=\sqrt{a_{11}/a_{22}}$,
we define the atom ratio $x$ as

	\begin{equation}
x=\frac {N_2} {N_1} \sqrt{\frac {a_{22}}{a_{11}}} \ ,
	\end{equation}
such that $x=1$ corresponds to the optimal mean-field composition. Note that 
one only needs to address the case $x\le 1$ due to the symmetric role of atom 
numbers.

	\section{\label{sec:stationary_drop_calc} Real-time relaxation of an isolated droplet}
	
	\begin{center}
	    \begin{figure}[t]
		\includegraphics[width=0.99\linewidth]{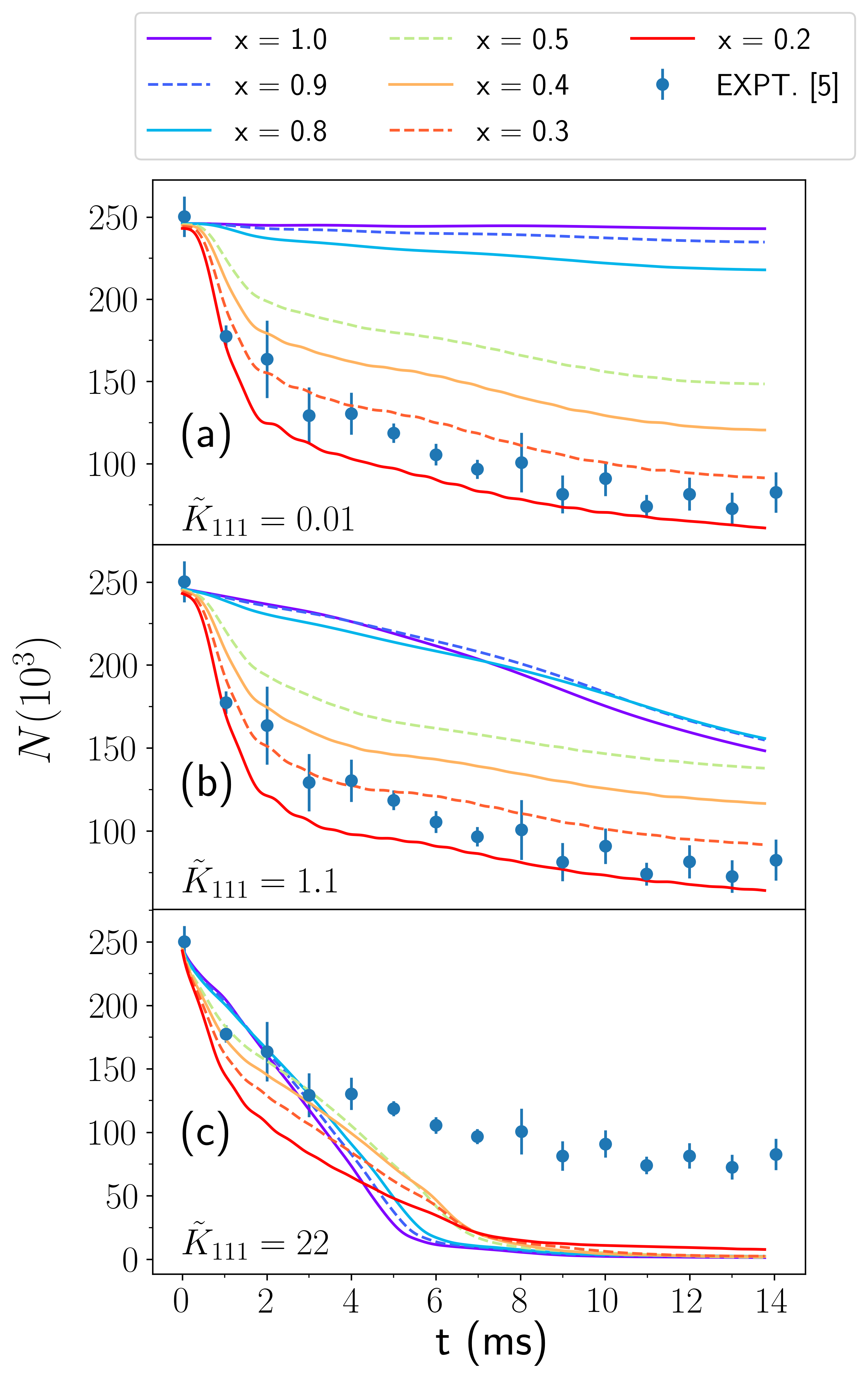}
		\caption{Time evolution of the total atom number $N$ in a stationary droplet. Lines are the results of two-component MF+LHY+3BL  calculations,
		 and points are experimental data from Ref. \cite{semeghini2018self}. The initial atom number is $N=2.5 \times 10^5$, with 
		 Gaussians of width $\sigma_r = 3\mu$m as density profiles. 		 
		 The values of $x=(N_2 / N_1) \sqrt{a_{22} / a_{11}}$ imposed at $t=0$ are given in  the top legend. 
		 The 3BL 
		 coefficients are shown  in the panels in units of $\hbar / (m\xi^2 \rho_0^2)$, with $\rho_0$ and $\xi$ being the equilibrium density and healing 
		 length given in Table \ref{table:scattering_parameters}.
		 }
		\label{fig:plotramp_A}
	\end{figure}
	\end{center}

	\begin{center}
	    \begin{figure}[t]
		\includegraphics[width=\linewidth]{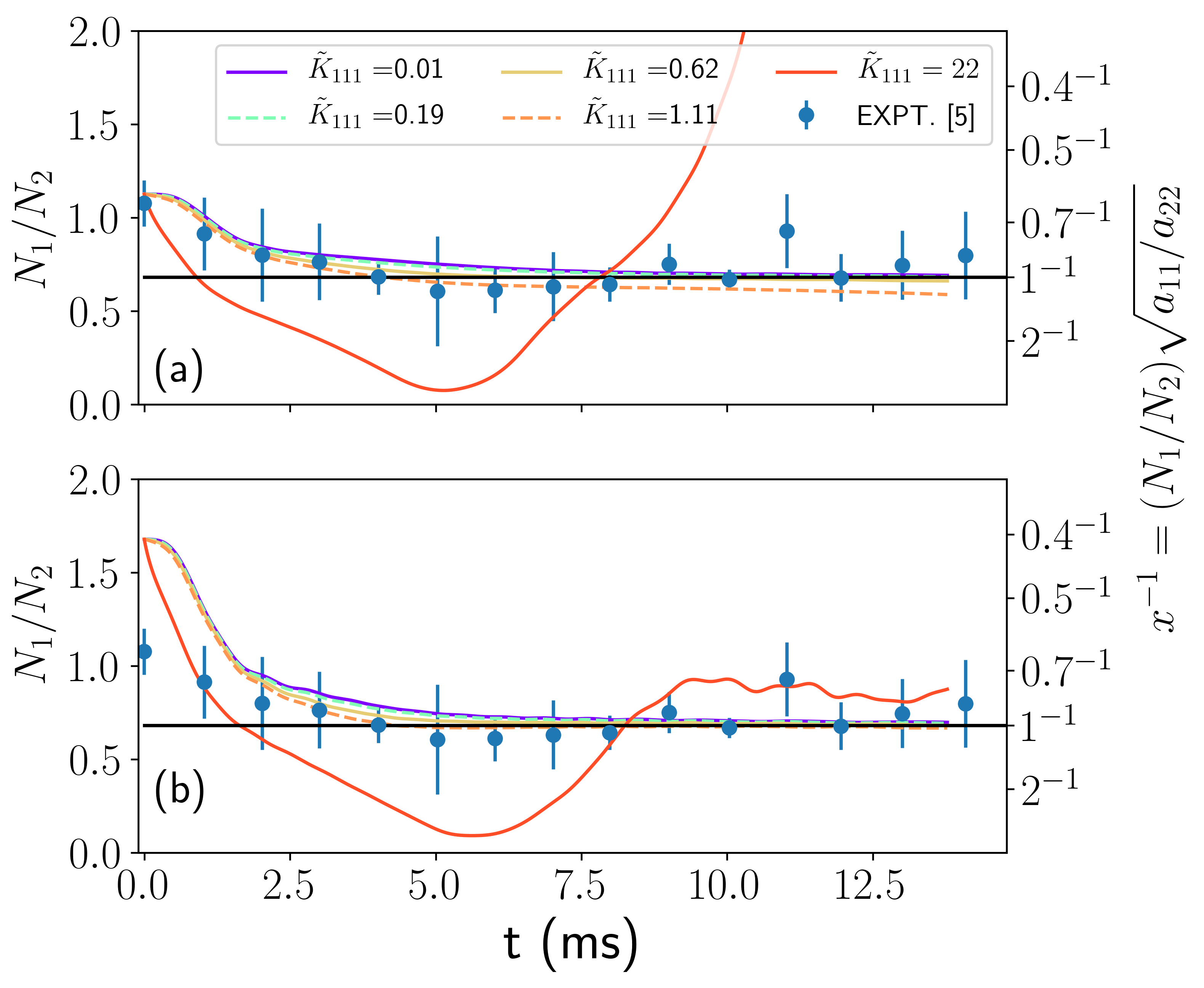}
		\caption{Time evolution of $N_1/N_2$ for the same simulations as in Fig. \ref{fig:plotramp_A}. 
		Starting atom number imbalances: panel (a), $x=0.7$; panel (b): $x=0.4$. 
		Points are data from the Ref. \cite{semeghini2018self}.
		}
		\label{fig:plotramp_B}
	\end{figure}
	\end{center}

	We first focus our attention on the time evolution of a 
$^{39}$K-$^{39}$K droplet as observed in the Florence experiment [Fig. (2) in Ref. 
\cite{semeghini2018self}]. We performed two-component MF+LHY calculations with 
the goal to investigate the influence of the initial non-optimal population 
ratio and 3BL on the real-time dynamics. 
%With the purpose of 
To mimic the experimental setup, at $t = 0$ we set 
the shape of a $N=2.5 \times 10^5$ atoms droplet as a Gaussian of width $\sigma_r = 3\,\mu$m 
%and composed by . 
and let the system evolve 
according to the extended Gross-Pitaevskii equations (\ref{gp1}) and 
(\ref{gp2}). The scattering parameters $a_{ij}$ correspond to the magnetic field 
$B = 56.54$ G (see Table \ref{table:scattering_parameters}). We denote the 
 $\ket{F, m_F} = \ket{1, 0}$ state as component 1 and the  $\ket{F, m_F} = 
\ket{1, -1}$ state as component 2. Within the two-component MF+LHY approach, we show in  Fig. \ref{fig:plotramp_A}  the time 
evolution of the total atom number, inside a volume of a cube with $ 12 \mu $m length centered in the origin of the simulation box,
for different initial atom ratios ranging from  $x=0.2$ to $x=1$. This was 
done 
%achieved numerically by 
changing the normalization of each component  to the desired value at $t=0$, 
keeping fixed the total number of atoms, $N=2.5 \times 10^5$. 
We have included 3BL only in the time evolution of the $\psi_1$ component, whose 3BL coefficient is reported to be at least 100 times larger than the 3BL coefficients in the other channels \cite{semeghini2018self,ferioli2020dynamical}.
This effect is included by introducing
a term $-i\hbar K_{111}|\psi_1|^4 / 2$ in Eq. (\ref{gp1}). We have explored three different 
values of $K_{111}$, namely $0$, $1.1$ 
and 22 in units of 
%\hbar / (m \xi^2\rho_0^2)$ and $22  
$\hbar / (m \xi^2\rho_0^2)$,  which correspond to 
$0$, $5.5 \times 10^{-28} \mathrm{~cm}^{6} / \mathrm{s}$ and 
$1.1 \times 10^{-26} \mathrm{~cm}^{6} / \mathrm{s}$, respectively (see Table \ref{table:scattering_parameters} for units).

Notice that there is a large experimental uncertainty 
affecting the actual value of $K_{111}$: both non-zero  values we
choose here are compatible with the experimental error bar in the
determination of the
3BL  term for the $^{39}$K-$^{39}$K mixture.
The largest of these $K_{111}$ values was
used in the two-component calculations in Ref. \cite{semeghini2018self} to explain the 
time evolution of the drop size observed in the experiment \cite{semeghini2018self}. %
Looking at  Fig. \ref{fig:plotramp_A}, it is clear that 
this value of $K_{111}$  
predicts too strong 3BL. For smaller  $K_{111}$ values (panels  (a) and (b) in Fig. \ref{fig:plotramp_A}), 
similar to those reported in  Ref. \cite{ferioli2019collisions}, we observe that 
compared to the  
atom number imbalance $x$, 3BL play a minor role in the equilibration process.
The relative atom 
number obtained from our calculations is also consistent with the experimental data as 
shown in Fig. \ref{fig:plotramp_B}, where  we display the evolution of relative 
atom number, starting from $x=0.7$ (panel a), and $x=0.4$ (panel b). 
The only exception is for the largest $K_{111}$ value which deviate significantly from the experimental results.
The observed atom loss can thus be mainly attributed to the droplet equilibration process 
in which the excess component is expelled  
until the optimal atom number ratio is eventually reached. A  much lesser contribution to the equilibration arises from 3BL.

	\begin{table}[t]
		\caption{Scattering parameters, i.e. $s$-wave scattering lengths $a$, in units of Bohr radius $a_0$, as a function of the magnetic field $B$ \cite{roy2013test}. $\xi$ and $\rho_0$ stand for healing length and equilibrium density of uniform liquid, respectively (see Eqs. \ref{eq:healing_length} and \ref{eq:rho_0}).}
		\label{table:scattering_parameters}
		\begin{tabular}{ c | c | c | c | c | c }
			\hline
			$B ({\rm G})$ & $a_{11} (a_0)$ & $a_{22} (a_0)$  & $a_{12} (a_0)$ & $\xi (\mathrm{\mu m})$ & $\rho_0 (\mathrm{cm}^{-3})$  \\
			\hline 
			56.23 & 63.648  & 34.587  & -53.435  &  0.496  & $5.773 \times 10^{15}$   \\
			56.54 &  72.960 & 33.962  & -53.293 & 1.479 & $1.217 \times 10^{15}$   \\
			56.55 &  73.3 & 33.9  & -53.3             &  1.527   & $1.164 \times 10^{15}$   \\
			%\hline 
		\end{tabular}
	\end{table}
	
\section{\label{sec:droplets_collisions}Droplets collisions}

We study the phase diagram characterizing the outcome of head-on binary 
collisions 
using the same description as in Ref. \cite{ferioli2019collisions}, i.e.,
in terms of $({v}, {N}_{\rm coll})$, where ${N}_{\rm coll} = {N}(t = t_{\rm coll})$ is the total atom number evaluated at the collision 
time $t_{\rm coll}$, and ${v}$ is velocity of each droplet at the beginning of the simulation. The collision 
time $t_{\rm coll}$ is estimated as $t_{\rm coll} = d /(2 v)$, where $d$ is the 
initial distance between the two droplets. The
initial wavefunction reads
\begin{equation}
\psi(t = 0)=\phi(x - d/2, y, z)e^{+ikx} + \phi(x + d/2, y, z)e^{-ikx},
\end{equation}
where $\phi$ and $k=mv/\hbar$ are the wave function and wave number of each droplet, respectively. Note that in all figures the total atom number $N$ and velocity $v$ are re-scaled according to Eqs. (\ref{eq:atom_number_mflhy}) and (\ref{eq:velocity_mflhy}). 

For all the simulations performed in this work we have observed 
three possible collision outcomes: 
(i) for small velocities, merging of the droplets into a single one 
(coalescence appears after substantial deformation of the droplets), 
(ii) for higher velocities, separation, where the two droplets move away from one another after the collision, 
and (iii) evaporation, which occurs for very energetic collisions. Shattering is not observed because BEC drops must have a minimum size to be bound \cite{petrov2015quantum} and instead of a cloud of small droplets the process continues until complete evaporation. 

\begin{center}
    \begin{figure}[t]
	\includegraphics[width=0.85\linewidth,keepaspectratio]{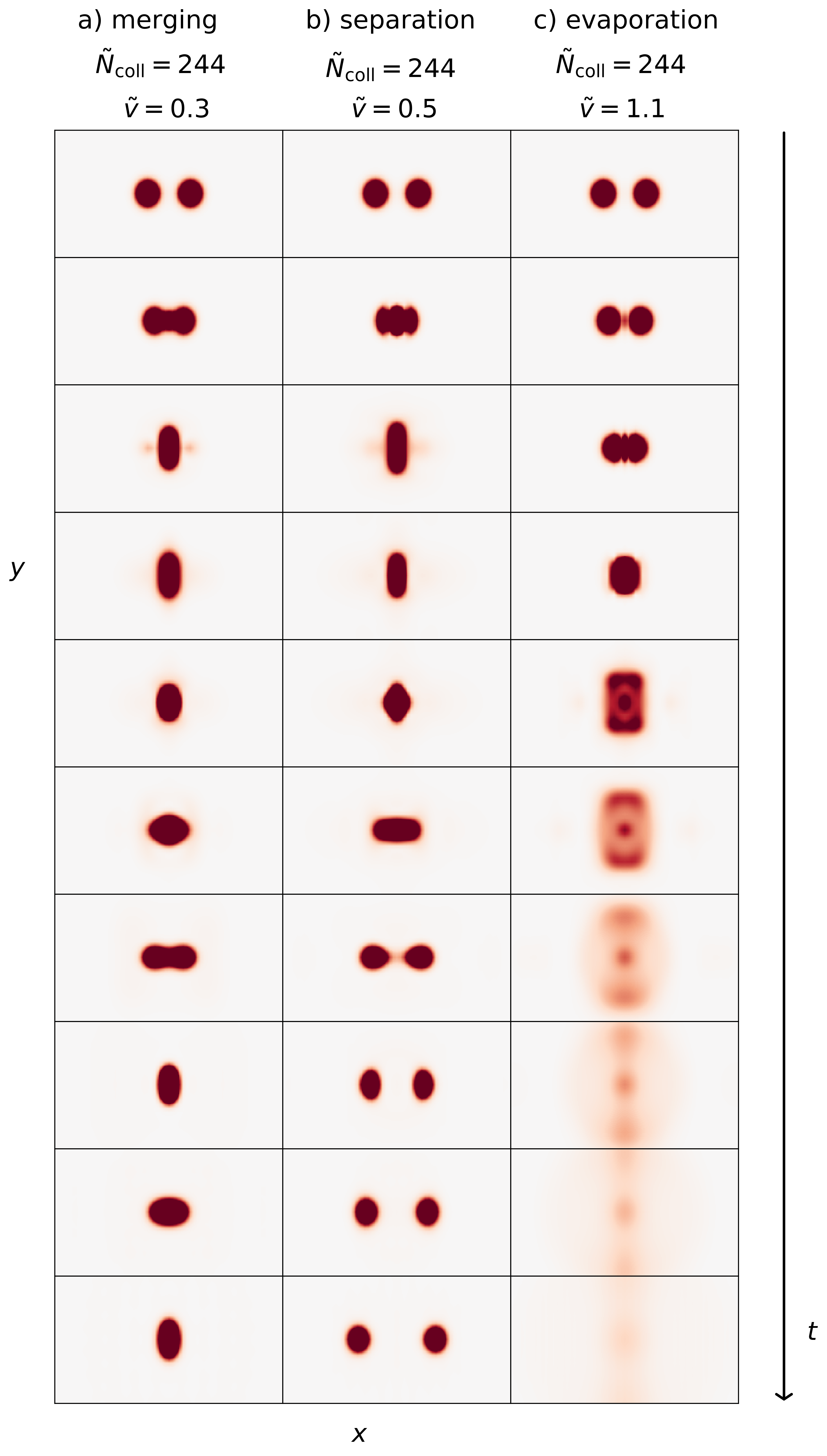}
	\caption{Three possible outcomes of a collision between quantum droplets 
		simulated by two-component MF+LHY calculations, without present three-body losses. 
		From left to right, the simulations correspond to $\tilde{v}=0.3$, $0.5$ and $1.1$ for $\tilde{N}_{\rm coll}=244$ in all three cases, at the magnetic field $B=56.55$ G.
		The panels show the integrated total atom density $\int dz \rho(\vec{r})$. The complete time-dependent evolution of the collisions is reported in Ref.~\cite{supplement}.		
	}		
	\label{fig:collisionoutcomes}
\end{figure}
\end{center}

This is illustrated  in Fig. \ref{fig:collisionoutcomes}, where one may see the three outcomes: merging (Fig. \ref{fig:collisionoutcomes}a),
separation (Fig. \ref{fig:collisionoutcomes}b), and evaporation (Fig. \ref{fig:collisionoutcomes}c). The simulation corresponds to the two-component calculations described below.

	\subsection{\label{sec:eff_single_comp_results} Effective single-component calculations}

\begin{center}
	    \begin{figure}[t]
		\includegraphics[width=\linewidth]{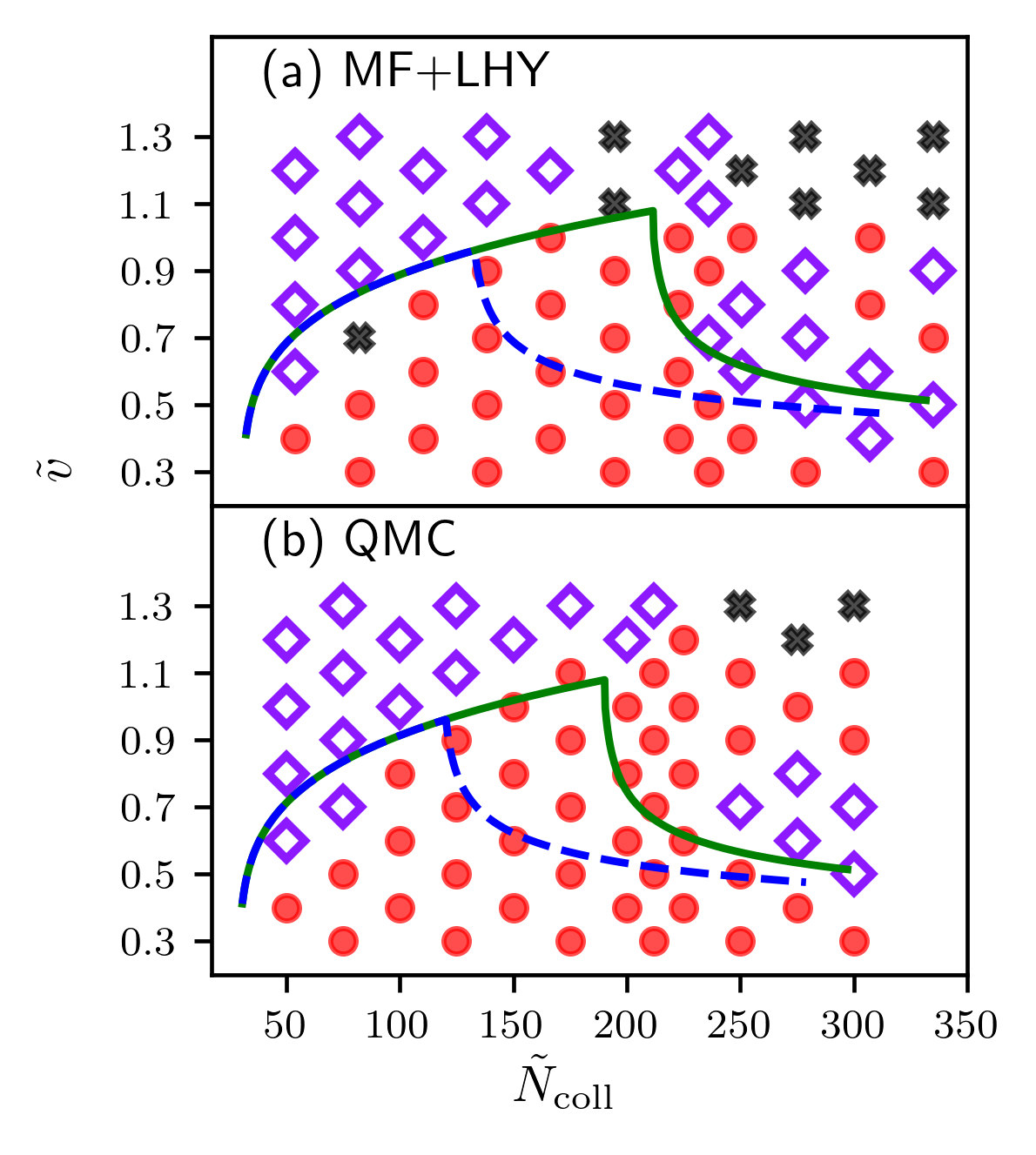}
		\caption{Collision outcomes obtained using the effective single-component MF+LHY 
		(a) and QMC functional (b) with the scattering 
		parameters corresponding to a magnetic field $B=56.23$ G (see Table 
\ref{table:scattering_parameters}). 
		3BL are not included. Atom number 
		$N$ and velocity of each droplet at $t=0$ are re-scaled according to Eqs. (\ref{eq:atom_number_mflhy}) and (\ref{eq:velocity_mflhy}), 
		respectively. Red dots, purple diamonds and black crosses stand for merging, separation and evaporation collision outcome 
		(see Fig. \ref{fig:collisionoutcomes}), respectively. Dashed blue and full green lines are the empirical fits to the experimental  and MF+LHY results~\cite{ferioli2019collisions}, respectively, for the velocity threshold above (below) which the droplets separate (merge). QMC functional is given by 
		Eq. (\ref{eq:single_comp_functional}) with parameters $\alpha = -0.812 \hbar^2  / (2m a_{11}^5)$, 
		$\beta = 5.974  \hbar^2 / (2m a_{11}^{5-3\gamma})$ and $\gamma = 1.276$, with $a_{11} = 63.648 a_0$. }
		\label{fig:plotvcnsinglecompk0mflhyqmc}
	\end{figure}
	\end{center}

    We have collected 
    in Fig. \ref{fig:plotvcnsinglecompk0mflhyqmc} 
    the collision outcomes obtained using the effective single-component MF+LHY and QMC functionals (Eq. (\ref{eq:single_comp_functional})), with the scattering parameters $a_{ij}$ corresponding to the magnetic field $B = 56.23$ G. Three-body losses are not  included. 
For small $N$, the two functionals do not yield significantly different collision outcomes. There are some differences for big droplets,
but not large enough to explain the deviation between theory and experiment observed in Ref. \cite{ferioli2019collisions}. 
Merging is more likely to occur when using this QMC functional, which is somewhat expected since 
such functional yields more binding \cite{cikojevic2020finite}
than the MF+LHY one, thus preventing the drops from separating.

From the results shown in Fig. \ref{fig:plotvcnsinglecompk0mflhyqmc} 
it appears that the effective single-component density functionals cannot account 
for the experimental observations since they predict droplets merging at much higher velocities than
experimentally observed. This is in agreement with the theoretical analysis in Ref. \cite{ferioli2019collisions}.

	\subsection{\label{sec:two_comp_calculations} Two-component calculations}

	We have performed collision simulations using the MF+LHY density functional in the two-component framework (Eqs. (\ref{gp1}) and (\ref{gp2})). Notice that this is not possible with the present QMC functional, as it is written in terms of the total density alone.  For each component, its wave function is evolved in time with the corresponding propagator (see Eqs. (\ref{eq:twocomp_equation_of_motion}) and (\ref{eq:time_evolution})).

\begin{center}
	    \begin{figure}[t]
		\includegraphics[width=\linewidth]{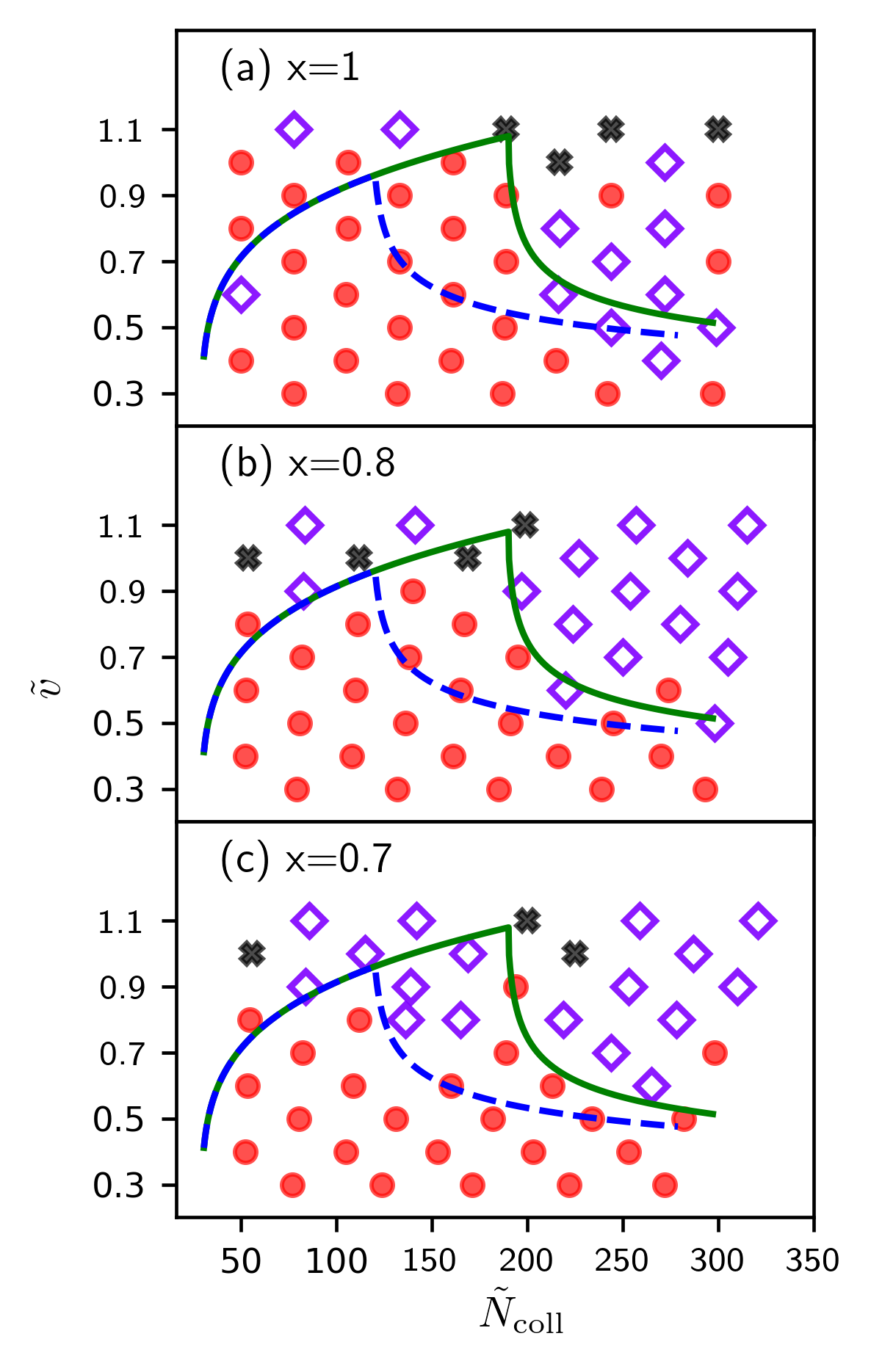}
		\caption{Collision outcome as a function of droplet velocity $\tilde{v}$ 
and total atom number evaluated at the instant of collision $\tilde{N}_{\rm 
coll}=\tilde{N}(t=t_{\rm coll})$ for the scattering parameters corresponding to 
$B=56.55$ G (see Table \ref{table:scattering_parameters}). Calculations are 
performed using two-component MF+LHY theory without including 3BL. Parameter 
$x= (N_2 / N_1) \sqrt{a_{22} / a_{11}}$ is the initial particle ratio at the 
beginning of collision. Points and lines have the same meaning as in Fig. 
\ref{fig:plotvcnsinglecompk0mflhyqmc}.}
		\label{fig:plotvcnk0twocompcombinedclosedistance}
	\end{figure}
	\end{center}

We summarize   in Fig. \ref{fig:plotvcnk0twocompcombinedclosedistance} the collision outcomes for several initial atom ratios $x$.
The functional we use is that obtained  from   
the scattering parameters $a_{ij}$ corresponding to the magnetic field $B=56.55$ G (see Table \ref{table:scattering_parameters}). 
In all cases the initial droplet separation is  
$d\approx 3500 \,a_{11} \approx 9 \,\xi$, where $\xi$ is 
the corresponding healing length (see Table \ref{table:scattering_parameters}). 
Since at $t=0$ the atom number ratio is not the optimal one,
i.e., it does not correspond to the ground-state of the extended 
Gross-Pitaevskii approach, 
the initial profile of each droplet has been prepared as follows. 
Firstly, each droplet is equilibrated with optimal atom composition $x = 1$ by 
means of imaginary-time propagation. Next, the 
real-time evolution is started and simultaneously the normalization of component 1 is changed to the
desired value of $x$. When the collision is started with $x = 1$ 
(Fig. \ref{fig:plotvcnk0twocompcombinedclosedistance}a), the predictions 
within the two-component framework are in good agreement with calculations 
from Ref. \cite{ferioli2019collisions} but in poor agreement with experiment. As we decrease $x$, our results 
for the critical velocity are in better accord with the experimental results. 

Note that there are small differences between the predictions obtained using effective single-component and two-component MF+LHY functionals assuming $x=1$. Since we are dealing with a finite system, the imposed requirement $\rho_2/\rho_1=\sqrt{a_{11}/a_{22}}$ satisfied in an effective single-component functional is not equally fulfilled in each spatial coordinate at the droplet surface in a two-component approach. Therefore, colliding drops display deviations of the density ratio $\rho_2/\rho_1$, which eventually leads to the difference in collision outcomes.

\begin{center}
	    \begin{figure}[t]
		\includegraphics[width=\linewidth]{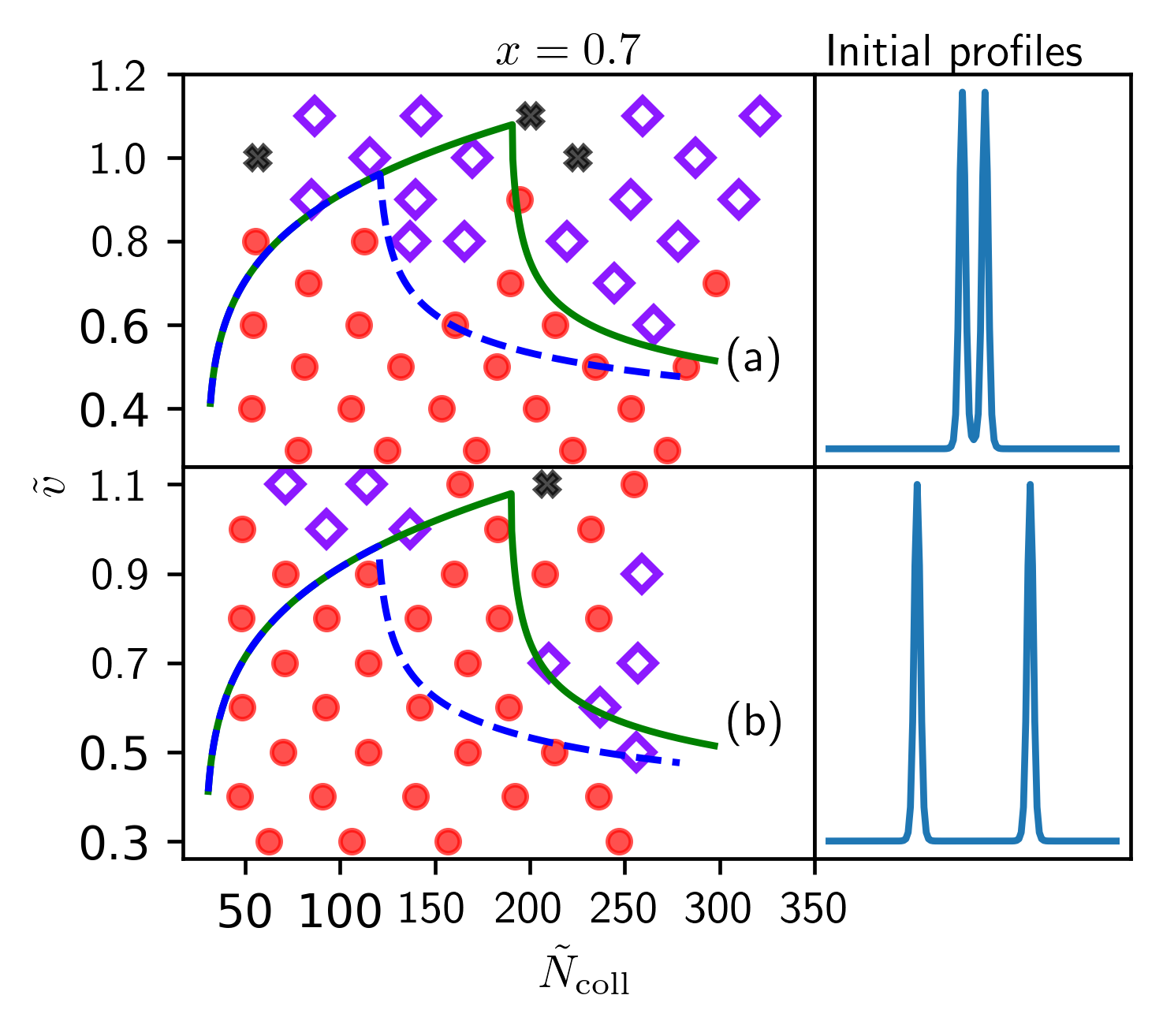}
		\caption{Same as Fig. \ref{fig:plotvcnk0twocompcombinedclosedistance} for the non-optimal atom number ratio $x=0.7$
		and two different initial droplets distances. Left: Predictions of merging (red points), 
		separation (purple diamonds) and evaporation (black crosses).
Right: initial droplets density profiles  along the approaching direction, $\rho(x, 0, 0; t=0)$. Full green and dashed blue lines have the same meaning as in Fig. \ref{fig:plotvcnsinglecompk0mflhyqmc}.
	Initial droplet distances in panel (a) are the same as in Fig. \ref{fig:plotvcnk0twocompcombinedclosedistance}c.
}
		\label{fig:diffinitdistancecombinedplot}
	\end{figure}
	\end{center}

	Since $x \neq 1$ is not the equilibrium configuration,  
the atoms of the in-excess component are
expelled out of the droplets as soon as time evolution starts.
Therefore, the initial configuration is somewhat ill-prepared and the collision outcome depends on the initial distance $d$
between the two droplets (the larger the $d$, the larger the amount of atoms
expelled from the droplets prior to  colliding). Moreover, the evaporated atoms do not leave the collision region and may also
influence the outcome.
To highlight this important 
effect, we report in Fig. \ref{fig:diffinitdistancecombinedplot}  the collision outcomes for non-equilibrated ($x=0.7$) droplets and
two different initial distances. Starting the collision at  a large distance $(d = 45 \xi)$,  
as in Fig. \ref{fig:diffinitdistancecombinedplot}b, the prediction for the 
critical velocity coincides with that obtained within 
the effective single-component framework. 
This is quite a natural result meaning that, by the time the two droplets meet, 
they have already reached a quasi-equilibrium configuration 
where the ratio of atoms in each component corresponds to $x = 1$.

	\subsection{\label{sec:three_body_losses} Effect of three-body losses and gas halo}
	
	We have also 
performed calculations including 3BL using the two-component MF+LHY density functional. 
Three-body recombination is assumed to be dominant in the 
$\ket{F, m_F} = \ket{1, 0}$ channel, which we call state 
$1$ \cite{ferioli2020dynamical, semeghini2018self}. 
Consequently, in our simulations 3BL act only on $\psi_1$, Eq. (\ref{gp1}).

\begin{center}
	    \begin{figure}[t]
		\includegraphics[width=\linewidth]{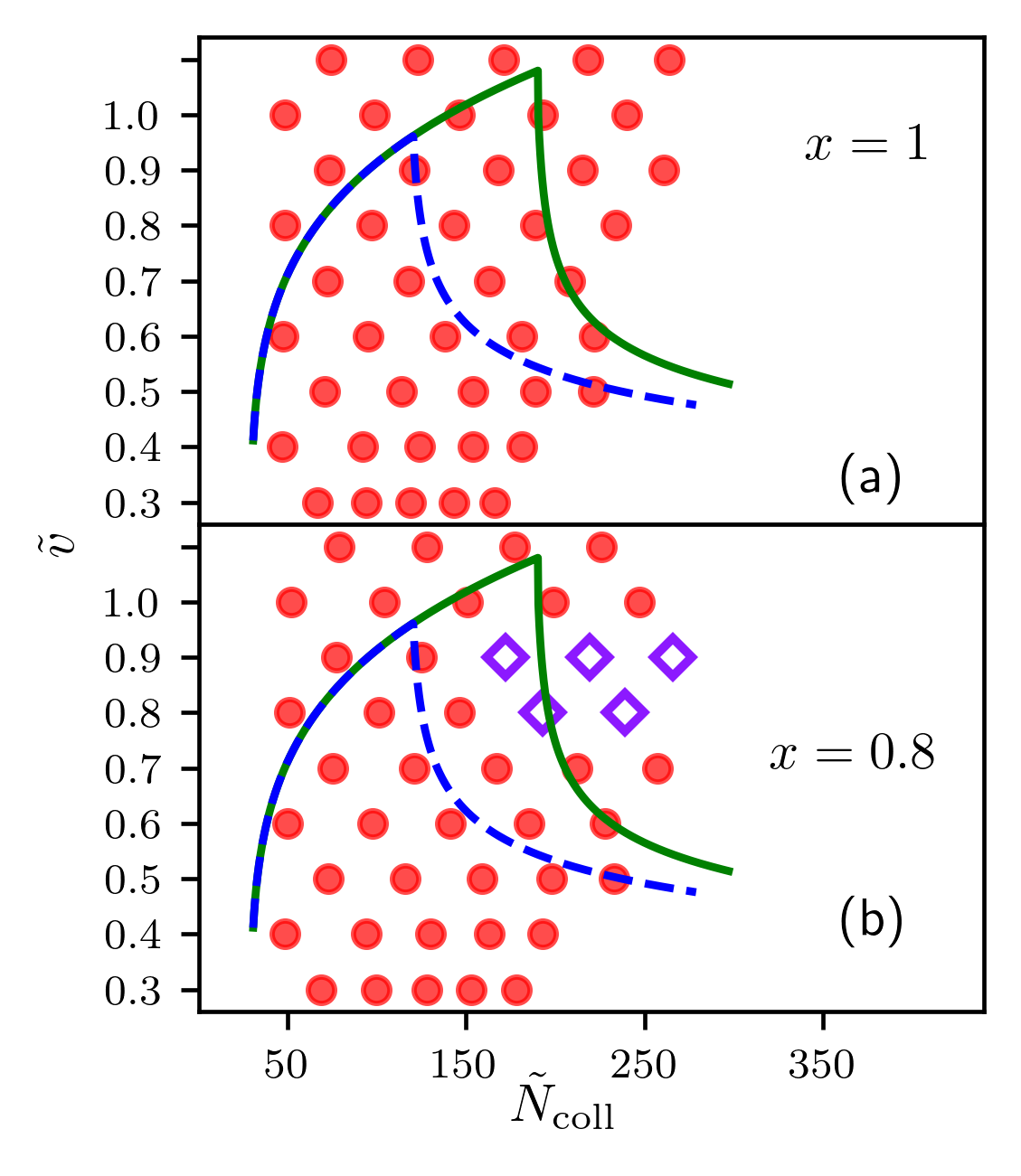}
		\caption{	Same as Fig. (\ref{fig:plotvcnk0twocompcombinedclosedistance}). Merging (red points) and separation (purple diamonds) 
		predicted within the two-component MF+LHY theory with 3BL included in the $\ket{F, m_F} = \ket{1, 0}$ component 
		using  $K_{111} = 2.73 \times 10^{-28} cm^6 / s$. 
		Upper figure shows the predictions when the initial atom number ratio is $x = (N_2/N_1) \sqrt{a_{22} / a_{11}} = 1$, 
		and the lower figure when $x = 0.8$.}
		\label{fig:combineplots3blonecomponent}
	\end{figure}
	\end{center}

We show in Fig. \ref{fig:combineplots3blonecomponent} the collision 
outcomes when the collision is started with the 
optimal atom number ratio $x = 1$ (Fig. \ref{fig:combineplots3blonecomponent}a) 
and with $x = 0.8$ (Fig. \ref{fig:combineplots3blonecomponent}b). 
The initial density profiles and the distance between the two droplets 
is the same as in Fig. \ref{fig:plotvcnk0twocompcombinedclosedistance}. 
The scattering parameters $a_{ij}$ correspond to the magnetic 
field $B=56.55$ G (see Table \ref{table:scattering_parameters}). 
We choose a value of $K_{111} = 2.73 \times 10^{-28} {\rm cm}^6 / {\rm s} = 
0.53 \hbar / (m \xi^2 \rho_0^2)$, i.e., 
the same as in Ref. \cite{ferioli2019collisions}, where it has 
been observed that the effective single-component theory, 
supplemented with a 3BL term $-i\hbar K\rho^2/2$, 
with $K=0.53 \hbar / (m \xi^2 \rho_0^2)$ and $\rho$ being the total atom density, 
allowed to reproduce the experimental curve dividing merging from separation.
Our calculations within two-component formalism show that only when the initial atom number ratio is non-optimal, 
namely $x = 0.8$, we observe separation-like outcomes of the collision, 
and only in part of the phase diagram, 
for droplet velocities  between $\tilde{v} = 0.7$ and $\tilde{v} = 1$.

We have also performed collisions using the effective single-component 
MF+LHY theory with the same value of three-body recombination 
as in Ref. \cite{ferioli2019collisions}, $\tilde{K}=0.53$, and we find that the 
collision outcome (and thus the overall aspect of the $v-N$ phase diagram) 
is very sensitive to the initial preparation of the droplets, namely the 
distance and the atom number at the start of the collision. In a majority of 
collisions performed, we have observed separation of the droplets, followed by 
the evaporation, even for  $(v, N)$ values for which merging is predicted in 
Ref. \cite{ferioli2019collisions}. Thus, only a precise knowledge of how the 
droplets have been prepared  (crucially, the initial droplet distance) would 
permit us to make a sensible comparison with the simulations of Ref. 
\cite{ferioli2019collisions}.

	The droplet dynamics becomes even more complex if we allow for the initial 
non-optimal atom ratio. The collision outcome sensibly depends on three key 
parameters: i) initial droplet distance, ii) initial atom ratio, and iii) the 
value of the three-body losses coefficient. When we use a non-optimal atom ratio, we 
observe that the expelled atoms of the in-excess species form a gas halo enveloping 
the colliding droplets, thus acting as a kind of weakly repulsive	buffer which 
slows the droplet relative velocity thus favoring merging over separation. Since 
the halo plays a role, we question the validity of the modeling of three-body 
losses since they just remove the atoms and energy from the simulation box.

\begin{figure*}
		\includegraphics[height=20 cm,keepaspectratio]{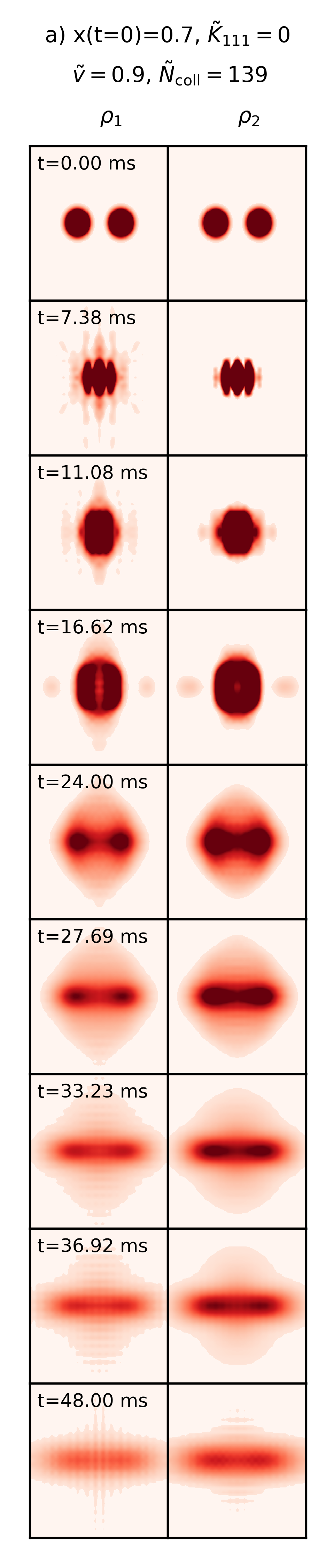}
		\includegraphics[height=20 cm,keepaspectratio]{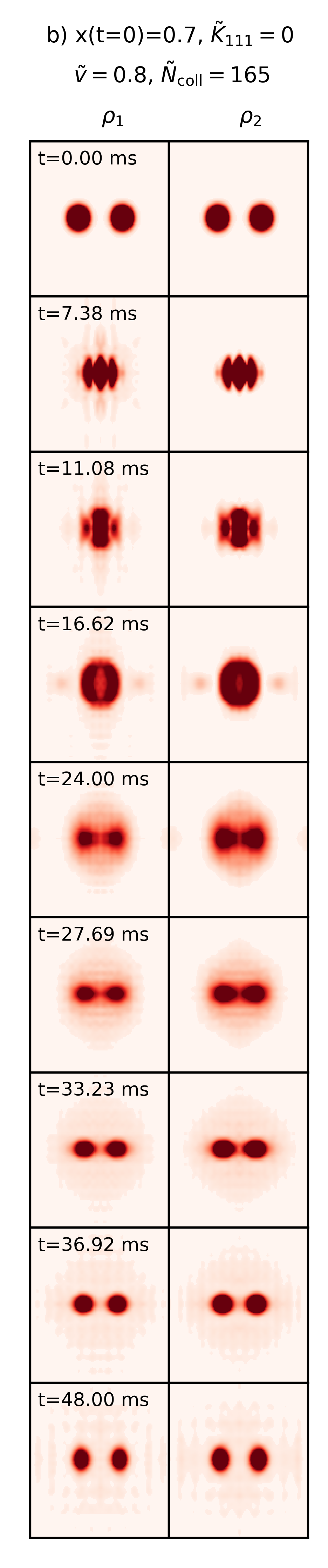}
		\includegraphics[height=20 cm,keepaspectratio]{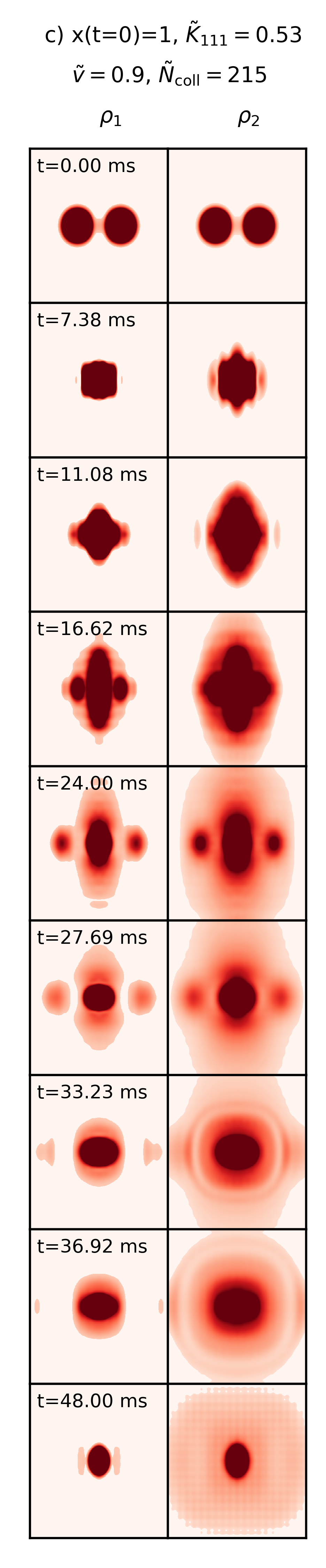}
		\includegraphics[height=20 cm,keepaspectratio]{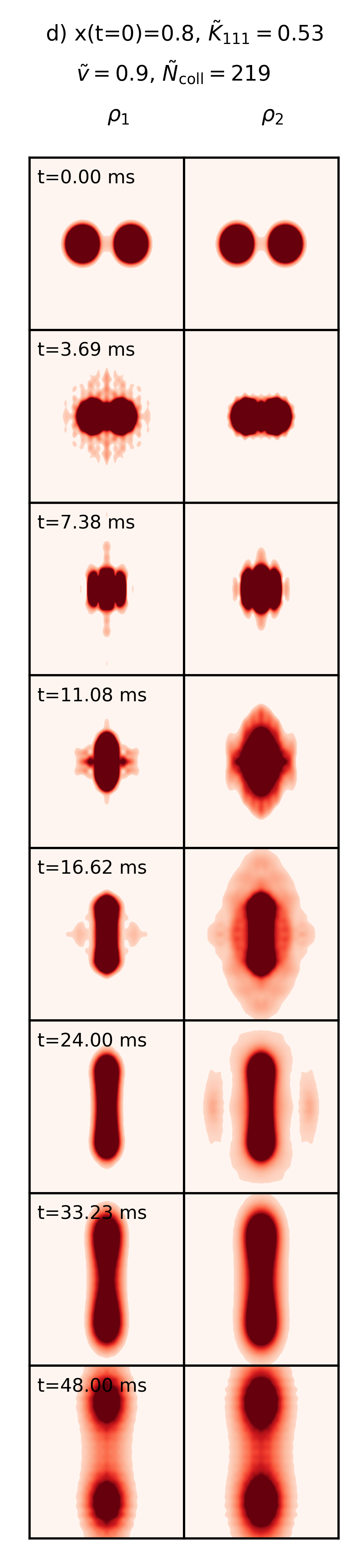}
		\caption{Images of the time evolution of $\rho_i=\int dz\rho_i(\vec{r})$, for components $i=1$ and $2$, obtained within the two-component MF+LHY functional. The complete time-dependent evolution of the collisions is reported in Ref.~\cite{supplement}. }
		\label{fig:collisionoutcomesx1no3bl}
	\end{figure*}

	To illustrate the effects of both the three-body losses and non-optimal atom 
ratios, we present in Fig. \ref{fig:collisionoutcomesx1no3bl} four types of 
collisions obtained within the two-component MF+LHY functional. In all four 
cases, the integrated density $\rho_i(x,y)=\int dz \rho_i(\vec{r})$ is shown, 
with component 1 (2) shown on the left (right).
	
	In Fig. \ref{fig:collisionoutcomesx1no3bl}a and 
\ref{fig:collisionoutcomesx1no3bl}b, 
the time-evolution is shown without three-body losses present in the system. 
The droplets are prepared with non-optimal atom ratio $x=0.7$, as in Fig. 
\ref{fig:plotvcnk0twocompcombinedclosedistance}. In both cases, separation is 
observed. For the smaller drops colliding at large velocity (Fig. 
\ref{fig:collisionoutcomesx1no3bl}a), drops evaporate upon separation since they 
require a minimum critical atom number to be self-bound 
\cite{petrov2015quantum}. When atom numbers are slightly higher (Fig. 
\ref{fig:collisionoutcomesx1no3bl}b), the remaining part of the drops separate. In both cases, there is a lot of evaporation in component 1 when the collision starts due 
to the initial population imbalance.

	In Fig. \ref{fig:collisionoutcomesx1no3bl}c and \ref{fig:collisionoutcomesx1no3bl}d we present the influence of the initial atom ratio on the collisions with the included three-body coefficient $\tilde{K}_{111}=0.53$, for initial atom ratio equal to $x=1$ and $x=0.8$, respectively. In this case, the interplay of three-body losses and atom number imbalance leads to different geometries during the collision, which eventually yield different collision outcomes. 
	When the initial atom ratio is $x=1$ (Fig. \ref{fig:collisionoutcomesx1no3bl}c), two protrusions are formed, reminiscent to collisions of He droplets  \cite{vicente2000coalescence,escartin2019vorticity}. The protrusions do not reach the size to be self-bound and eventually evaporate. 
	
	During the whole process, and more pronounced during the collision, when the central densities increase a lot, normalization of density 1 drops due to 3BL. This is followed by evaporation in component 2 by a mechanism of equilibration to optimal particle number, which creates a halo around the droplets. Around $t=30$ ms, a shock wave is formed in component 2, as a result of the interference of outward expansion of the droplet with the surrounding halo.
	Eventually, the shock wave passes through the halo, and the drops remain at place. This result highlights that evaporated atoms that remain around the colliding droplets contribute to the merging process. Indeed, it has been experimentally found that coalescence of 	viscous droplets can be facilitated when the collision region contains atoms in the gas phase \cite{qian1997regimes}. Notice that in order for the droplets to merge, the gas between them must be expelled, which costs some energy favoring the merging.

	When the collision is started with $x=0.8$ (Fig. \ref{fig:collisionoutcomesx1no3bl}d), evaporation of component 1 in the early stage of collision takes place. This happens due to equilibration towards optimal particle composition, which would happen even in the absence of 3BL (see Fig. \ref{fig:plotramp_A}). At about $t = 25$ ms a different geometry and a thinner halo of species 2 atoms than in the case of $x(t=0)=1$ collision (Fig. \ref{fig:collisionoutcomesx1no3bl}c) is formed. This is due to the fact that in a time window when $x<1$, only the component 1 evaporates, not the component 2. The drops form a peanut-like configuration which has a much longer timelife, also observed in \cite{ferioli2019collisions}, during which the large part of component 2 halo evaporates away. The peanut/cylinder eventually splits into two  pieces due to surface tension. All of the separation collision outcomes observed in Fig. \ref{fig:combineplots3blonecomponent}b are of the kind displayed in Fig. \ref{fig:collisionoutcomesx1no3bl}d, indicating that this could be a common feature for collisions with initial population imbalance and three-body losses present only in channel 111.

	\section{\label{sec:conclusions} Summary and conclusions}

	The recent study of head-on $^{39}$K-$^{39}$K droplet collisions 
\cite{ferioli2019collisions} offers a new avenue of research by extending the 
study of quantum droplet collisions -- previously restricted to the case of 
helium droplets \cite{barranco2006helium} to much lower density and temperature 
ranges of ultra-dilute cold Bose gas mixtures. In the present work, we have 
theoretically reanalyzed this experiment by introducing elements not considered 
in the original work. In particular, we have improved the density functional 
approach by considering a functional based on QMC calculations that correctly 
incorporates the two relevant scattering parameters of the $^{39}$K-$^{39}$K 
mixture, namely the $s$-wave scattering length and the effective range, as well 
as the most general version of a two-component equal-masses Bose-Bose energy 
functional. This two-component functional allows to introduce the 3BL only in 
the most affected component of the mixture, instead of in the total density of 
the system. This is a crucial improvement over the effective single-component 
functionals like QMC-based one, or the effective single-component MF+LHY 
functional which follows from the requirement that the two components of the 
mixture have the density ratio corresponding to the equilibrium one (the 
approach used in 
Ref. \cite{ferioli2019collisions}).

	Our results can be summarized as follows:
	
	(1) When 3BL are not considered, neither the QMC nor the effective single 
component MF+LHY approaches agree with experiment. As already noted in the 
theoretical analysis of Ref. \cite{ferioli2019collisions}, the MF+LHY approach 
yields droplet merging at much higher velocities than observed experimentally. 
Moreover, the QMC-based and the single-component MF+LHY functionals meaningfully 
yield similar results. This is quite surprising since the QMC-based functional 
has a rather larger incompressibility and binding energy per atom than the 
single-component MF+LHY functional \cite{cikojevic2020finite}, which was the 
reason to use it in this study. 
	
	(2) Given the experimental findings on the intensity of the 3BL in each component of the mixture \cite{semeghini2018self,ferioli2020dynamical}, it is not justified to let the 3BL act on both components, since it is about 100 times more effective in the component $\ket{F, m_F} = \ket{1, 0}$ than in the other. There is little justification for using a single-component MF+LHY approach. When the value of the $K_{111}$ coefficient is in the range of those compatible with the experiments \cite{ferioli2019collisions}, we find poor agreement with the experimental collision results. This implies that 3BL alone acting on equilibrated droplets cannot resolve the disagreement between the experiment and the results of either the effective single-component or the two-component MF+LHY approach, and we have been prompted to explore other natural possibilities.
	
	(3) One of the possibilities is that droplets are not fully equilibrated when the collision is considered to have started. We completed our analysis assuming that the droplets are not in equilibrium at the start of the collision, including in some cases 3BL, always only in the component $\ket{F, m_F} = \ket{1, 0}$. 
	
	If the droplets are not fully equilibrated when they enter the collision region, this affects the collision outcome because it introduces an important additional effect not previously considered, namely the halo of the expelled gaseous particles of the in-excess species that envelop the droplets and increase the tendency to merge. This effect has also been found in viscous droplet collisions \cite{qian1997regimes}. Remarkably, the phase diagram changes even without the introduction  of 3BL, as collisions between non-equilibrated droplets are shown to behave differently from equilibrated ones in terms of the optimal atom number ratio, leading to results that are in better agreement with experiment. We have focused here on zero temperature description, but it is plausible that thermally excited droplets \cite{wang2020thermal} could produce similar shifts in the critical velocity. 
	
	It is evident that the introduction of considerations of 3BL and/or non-equilibrium configuration at $t=0$ in the description of the collision process have a dramatic impact on the outcome of the simulations and add elements of difficult control when comparing with experiment.
	
	Finally, we want to emphasize that while 3BL and non-equilibrium effects reduce the number of atoms and the kinetic energy in the colliding droplets, both effects are by no means equivalent. The term added to the functional to introduce 3BL takes atoms and energy out of the computational box, while the atoms of the in-excess component remain in the collision region and their presence may affect the collision outcome.

	 \acknowledgments

        This work has been supported by the Ministerio de Economia, Industria y Competitividad (MINECO, Spain) under grants
        Nos. FIS2017-84114-C2-1-P and FIS2017-87801-P (AEI/FEDER, UE), and by the EC Research Innovation Action under the H2020 Programme,
        Project HPC-EUROPA3 (INFRAIA-2016-1-730897). V. C. gratefully acknowledges the
        computer resources and technical support provided by Barcelona Supercomputing Center.
        We also acknowledge financial support from Secretaria d'Universitats i Recerca del Departament d'Empresa i Coneixement de la Generalitat de Catalunya, co-funded by the European Union Regional Development Fund within the ERDF Operational Program of Catalunya (project QuantumCat, ref. 001-P-001644).

	\bibliography{paper_collisions}% Produces the bibliography via BibTeX.

\begin{thebibliography}{63}%
\makeatletter
\providecommand \@ifxundefined [1]{%
 \@ifx{#1\undefined}
}%
\providecommand \@ifnum [1]{%
 \ifnum #1\expandafter \@firstoftwo
 \else \expandafter \@secondoftwo
 \fi
}%
\providecommand \@ifx [1]{%
 \ifx #1\expandafter \@firstoftwo
 \else \expandafter \@secondoftwo
 \fi
}%
\providecommand \natexlab [1]{#1}%
\providecommand \enquote  [1]{``#1''}%
\providecommand \bibnamefont  [1]{#1}%
\providecommand \bibfnamefont [1]{#1}%
\providecommand \citenamefont [1]{#1}%
\providecommand \href@noop [0]{\@secondoftwo}%
\providecommand \href [0]{\begingroup \@sanitize@url \@href}%
\providecommand \@href[1]{\@@startlink{#1}\@@href}%
\providecommand \@@href[1]{\endgroup#1\@@endlink}%
\providecommand \@sanitize@url [0]{\catcode `\\12\catcode `\$12\catcode
  `\&12\catcode `\#12\catcode `\^12\catcode `\_12\catcode `\%12\relax}%
\providecommand \@@startlink[1]{}%
\providecommand \@@endlink[0]{}%
\providecommand \url  [0]{\begingroup\@sanitize@url \@url }%
\providecommand \@url [1]{\endgroup\@href {#1}{\urlprefix }}%
\providecommand \urlprefix  [0]{URL }%
\providecommand \Eprint [0]{\href }%
\providecommand \doibase [0]{http://dx.doi.org/}%
\providecommand \selectlanguage [0]{\@gobble}%
\providecommand \bibinfo  [0]{\@secondoftwo}%
\providecommand \bibfield  [0]{\@secondoftwo}%
\providecommand \translation [1]{[#1]}%
\providecommand \BibitemOpen [0]{}%
\providecommand \bibitemStop [0]{}%
\providecommand \bibitemNoStop [0]{.\EOS\space}%
\providecommand \EOS [0]{\spacefactor3000\relax}%
\providecommand \BibitemShut  [1]{\csname bibitem#1\endcsname}%
\let\auto@bib@innerbib\@empty
%</preamble>
\bibitem [{\citenamefont {Ashgriz}\ and\ \citenamefont
  {Poo}(1990)}]{ashgriz1990coalescence}%
  \BibitemOpen
  \bibfield  {author} {\bibinfo {author} {\bibfnamefont {N.}~\bibnamefont
  {Ashgriz}}\ and\ \bibinfo {author} {\bibfnamefont {J.~Y.}\ \bibnamefont
  {Poo}},\ }\href@noop {} {\bibfield  {journal} {\bibinfo  {journal} {Journal
  of Fluid Mechanics}\ }\textbf {\bibinfo {volume} {221}},\ \bibinfo {pages}
  {183–204} (\bibinfo {year} {1990})}\BibitemShut {NoStop}%
\bibitem [{\citenamefont {Qian}\ and\ \citenamefont
  {Law}(1997)}]{qian1997regimes}%
  \BibitemOpen
  \bibfield  {author} {\bibinfo {author} {\bibfnamefont {J.}~\bibnamefont
  {Qian}}\ and\ \bibinfo {author} {\bibfnamefont {C.~K.}\ \bibnamefont {Law}},\
  }\href@noop {} {\bibfield  {journal} {\bibinfo  {journal} {Journal of Fluid
  Mechanics}\ }\textbf {\bibinfo {volume} {331}},\ \bibinfo {pages} {59–80}
  (\bibinfo {year} {1997})}\BibitemShut {NoStop}%
\bibitem [{\citenamefont {Pan}\ \emph {et~al.}(2009)\citenamefont {Pan},
  \citenamefont {Chou},\ and\ \citenamefont {Tseng}}]{Pan09}%
  \BibitemOpen
  \bibfield  {author} {\bibinfo {author} {\bibfnamefont {K.-L.}\ \bibnamefont
  {Pan}}, \bibinfo {author} {\bibfnamefont {P.-C.}\ \bibnamefont {Chou}}, \
  and\ \bibinfo {author} {\bibfnamefont {Y.-J.}\ \bibnamefont {Tseng}},\
  }\href@noop {} {\bibfield  {journal} {\bibinfo  {journal} {Phys. Rev. E}\
  }\textbf {\bibinfo {volume} {80}},\ \bibinfo {pages} {036301} (\bibinfo
  {year} {2009})}\BibitemShut {NoStop}%
\bibitem [{\citenamefont {Varma}\ \emph {et~al.}(2016)\citenamefont {Varma},
  \citenamefont {Ray}, \citenamefont {Wang}, \citenamefont {Wang},\ and\
  \citenamefont {Ramanujan}}]{Var16}%
  \BibitemOpen
  \bibfield  {author} {\bibinfo {author} {\bibfnamefont {V.}~\bibnamefont
  {Varma}}, \bibinfo {author} {\bibfnamefont {A.}~\bibnamefont {Ray}}, \bibinfo
  {author} {\bibfnamefont {Z.}~\bibnamefont {Wang}}, \bibinfo {author}
  {\bibfnamefont {Z.}~\bibnamefont {Wang}}, \ and\ \bibinfo {author}
  {\bibfnamefont {R.}~\bibnamefont {Ramanujan}},\ }\href@noop {} {\bibfield
  {journal} {\bibinfo  {journal} {Scientific reports}\ }\textbf {\bibinfo
  {volume} {6}},\ \bibinfo {pages} {1} (\bibinfo {year} {2016})}\BibitemShut
  {NoStop}%
\bibitem [{\citenamefont {Swiatecki}(1982)}]{Swi82}%
  \BibitemOpen
  \bibfield  {author} {\bibinfo {author} {\bibfnamefont {W.~J.}\ \bibnamefont
  {Swiatecki}},\ }\href@noop {} {\bibfield  {journal} {\bibinfo  {journal}
  {Nuclear Physics A}\ }\textbf {\bibinfo {volume} {376}},\ \bibinfo {pages}
  {275} (\bibinfo {year} {1982})}\BibitemShut {NoStop}%
\bibitem [{\citenamefont {Guilleumas}\ \emph {et~al.}(1995)\citenamefont
  {Guilleumas}, \citenamefont {Pi}, \citenamefont {Barranco},\ and\
  \citenamefont {Suraud}}]{Gui95}%
  \BibitemOpen
  \bibfield  {author} {\bibinfo {author} {\bibfnamefont {M.}~\bibnamefont
  {Guilleumas}}, \bibinfo {author} {\bibfnamefont {M.}~\bibnamefont {Pi}},
  \bibinfo {author} {\bibfnamefont {M.}~\bibnamefont {Barranco}}, \ and\
  \bibinfo {author} {\bibfnamefont {E.}~\bibnamefont {Suraud}},\ }\href@noop {}
  {\bibfield  {journal} {\bibinfo  {journal} {Zeitschrift f{\"u}r Physik D
  Atoms, Molecules and Clusters}\ }\textbf {\bibinfo {volume} {34}},\ \bibinfo
  {pages} {35} (\bibinfo {year} {1995})}\BibitemShut {NoStop}%
\bibitem [{\citenamefont {Toennies}\ and\ \citenamefont
  {Vilesov}(2004)}]{toennies2004superfluid}%
  \BibitemOpen
  \bibfield  {author} {\bibinfo {author} {\bibfnamefont {J.~P.}\ \bibnamefont
  {Toennies}}\ and\ \bibinfo {author} {\bibfnamefont {A.~F.}\ \bibnamefont
  {Vilesov}},\ }\href@noop {} {\bibfield  {journal} {\bibinfo  {journal}
  {Angewandte Chemie International Edition}\ }\textbf {\bibinfo {volume}
  {43}},\ \bibinfo {pages} {2622} (\bibinfo {year} {2004})}\BibitemShut
  {NoStop}%
\bibitem [{\citenamefont {Donnelly}(1993)}]{Don91}%
  \BibitemOpen
  \bibfield  {author} {\bibinfo {author} {\bibfnamefont {R.~J.}\ \bibnamefont
  {Donnelly}},\ }\href@noop {} {\emph {\bibinfo {title} {Quantized vortices in
  helium II}}},\ Vol.~\bibinfo {volume} {3}\ (\bibinfo  {publisher} {Cambridge
  Studies in Low Temperature Physics, Cambridge University Press, Cambridge,
  U.K.},\ \bibinfo {year} {1993})\BibitemShut {NoStop}%
\bibitem [{\citenamefont {Tsubota}\ and\ \citenamefont
  {Halperin}(2009)}]{Tsu09}%
  \BibitemOpen
  \bibfield  {author} {\bibinfo {author} {\bibfnamefont {M.}~\bibnamefont
  {Tsubota}}\ and\ \bibinfo {author} {\bibfnamefont {W.~P.}\ \bibnamefont
  {Halperin}},\ }\href@noop {} {\emph {\bibinfo {title} {Bose-Einstein
  condensation and superfluidity}}},\ Vol.\ \bibinfo {volume} {XVI}\ (\bibinfo
  {publisher} {Progress in Low Temperature Physics},\ \bibinfo {year}
  {2009})\BibitemShut {NoStop}%
\bibitem [{\citenamefont {Vicente}\ \emph {et~al.}(2000)\citenamefont
  {Vicente}, \citenamefont {Kim}, \citenamefont {Maris},\ and\ \citenamefont
  {Seidel}}]{vicente2000coalescence}%
  \BibitemOpen
  \bibfield  {author} {\bibinfo {author} {\bibfnamefont {C.}~\bibnamefont
  {Vicente}}, \bibinfo {author} {\bibfnamefont {C.}~\bibnamefont {Kim}},
  \bibinfo {author} {\bibfnamefont {H.}~\bibnamefont {Maris}}, \ and\ \bibinfo
  {author} {\bibfnamefont {G.}~\bibnamefont {Seidel}},\ }\href@noop {}
  {\bibfield  {journal} {\bibinfo  {journal} {Journal of low temperature
  physics}\ }\textbf {\bibinfo {volume} {121}},\ \bibinfo {pages} {627}
  (\bibinfo {year} {2000})}\BibitemShut {NoStop}%
\bibitem [{\citenamefont {Maris}(2003)}]{maris2003analysis}%
  \BibitemOpen
  \bibfield  {author} {\bibinfo {author} {\bibfnamefont {H.~J.}\ \bibnamefont
  {Maris}},\ }\href@noop {} {\bibfield  {journal} {\bibinfo  {journal}
  {Physical Review E}\ }\textbf {\bibinfo {volume} {67}},\ \bibinfo {pages}
  {066309} (\bibinfo {year} {2003})}\BibitemShut {NoStop}%
\bibitem [{\citenamefont {Ishiguro}\ \emph {et~al.}(2004)\citenamefont
  {Ishiguro}, \citenamefont {Graner}, \citenamefont {Rolley},\ and\
  \citenamefont {Balibar}}]{ishiguro2004coalescence}%
  \BibitemOpen
  \bibfield  {author} {\bibinfo {author} {\bibfnamefont {R.}~\bibnamefont
  {Ishiguro}}, \bibinfo {author} {\bibfnamefont {F.}~\bibnamefont {Graner}},
  \bibinfo {author} {\bibfnamefont {E.}~\bibnamefont {Rolley}}, \ and\ \bibinfo
  {author} {\bibfnamefont {S.}~\bibnamefont {Balibar}},\ }\href@noop {}
  {\bibfield  {journal} {\bibinfo  {journal} {Phys. Rev. Lett.}\ }\textbf
  {\bibinfo {volume} {93}},\ \bibinfo {pages} {235301} (\bibinfo {year}
  {2004})}\BibitemShut {NoStop}%
\bibitem [{\citenamefont {Escart\'{\i}n}\ \emph {et~al.}(2019)\citenamefont
  {Escart\'{\i}n}, \citenamefont {Ancilotto}, \citenamefont {Barranco},\ and\
  \citenamefont {Pi}}]{escartin2019vorticity}%
  \BibitemOpen
  \bibfield  {author} {\bibinfo {author} {\bibfnamefont {J.~M.}\ \bibnamefont
  {Escart\'{\i}n}}, \bibinfo {author} {\bibfnamefont {F.}~\bibnamefont
  {Ancilotto}}, \bibinfo {author} {\bibfnamefont {M.}~\bibnamefont {Barranco}},
  \ and\ \bibinfo {author} {\bibfnamefont {M.}~\bibnamefont {Pi}},\ }\href@noop
  {} {\bibfield  {journal} {\bibinfo  {journal} {Phys. Rev. B}\ }\textbf
  {\bibinfo {volume} {99}},\ \bibinfo {pages} {140505} (\bibinfo {year}
  {2019})}\BibitemShut {NoStop}%
\bibitem [{\citenamefont {Nazarenko}(2015)}]{naz15}%
  \BibitemOpen
  \bibfield  {author} {\bibinfo {author} {\bibfnamefont {S.}~\bibnamefont
  {Nazarenko}},\ }\href@noop {} {\bibfield  {journal} {\bibinfo  {journal}
  {Contemporary Physics}\ }\textbf {\bibinfo {volume} {56}},\ \bibinfo {pages}
  {359} (\bibinfo {year} {2015})}\BibitemShut {NoStop}%
\bibitem [{\citenamefont {Sun}\ and\ \citenamefont
  {Pindzola}(2008)}]{sun2008slow}%
  \BibitemOpen
  \bibfield  {author} {\bibinfo {author} {\bibfnamefont {B.}~\bibnamefont
  {Sun}}\ and\ \bibinfo {author} {\bibfnamefont {M.}~\bibnamefont {Pindzola}},\
  }\href@noop {} {\bibfield  {journal} {\bibinfo  {journal} {Journal of Physics
  B: Atomic, Molecular and Optical Physics}\ }\textbf {\bibinfo {volume}
  {41}},\ \bibinfo {pages} {155302} (\bibinfo {year} {2008})}\BibitemShut
  {NoStop}%
\bibitem [{\citenamefont {Astrakharchik}\ and\ \citenamefont
  {Malomed}(2018)}]{astrakharchik2018dynamics}%
  \BibitemOpen
  \bibfield  {author} {\bibinfo {author} {\bibfnamefont {G.~E.}\ \bibnamefont
  {Astrakharchik}}\ and\ \bibinfo {author} {\bibfnamefont {B.~A.}\ \bibnamefont
  {Malomed}},\ }\href@noop {} {\bibfield  {journal} {\bibinfo  {journal} {Phys.
  Rev. A}\ }\textbf {\bibinfo {volume} {98}},\ \bibinfo {pages} {013631}
  (\bibinfo {year} {2018})}\BibitemShut {NoStop}%
\bibitem [{\citenamefont {Zhong}(2021)}]{zhong2021oscillatory}%
  \BibitemOpen
  \bibfield  {author} {\bibinfo {author} {\bibfnamefont {H.}~\bibnamefont
  {Zhong}},\ }\href@noop {} {\bibfield  {journal} {\bibinfo  {journal}
  {Communications in Theoretical Physics}\ } (\bibinfo {year}
  {2021})}\BibitemShut {NoStop}%
\bibitem [{\citenamefont {Pitaevskii}\ and\ \citenamefont
  {Stringari}(2016)}]{pitaevskii2016bose}%
  \BibitemOpen
  \bibfield  {author} {\bibinfo {author} {\bibfnamefont {L.}~\bibnamefont
  {Pitaevskii}}\ and\ \bibinfo {author} {\bibfnamefont {S.}~\bibnamefont
  {Stringari}},\ }\href@noop {} {\emph {\bibinfo {title} {Bose-Einstein
  condensation and superfluidity}}},\ Vol.\ \bibinfo {volume} {164}\ (\bibinfo
  {publisher} {Oxford University Press},\ \bibinfo {year} {2016})\BibitemShut
  {NoStop}%
\bibitem [{\citenamefont {Dalfovo}\ \emph {et~al.}(1999)\citenamefont
  {Dalfovo}, \citenamefont {Giorgini}, \citenamefont {Pitaevskii},\ and\
  \citenamefont {Stringari}}]{dalfovo1999theory}%
  \BibitemOpen
  \bibfield  {author} {\bibinfo {author} {\bibfnamefont {F.}~\bibnamefont
  {Dalfovo}}, \bibinfo {author} {\bibfnamefont {S.}~\bibnamefont {Giorgini}},
  \bibinfo {author} {\bibfnamefont {L.~P.}\ \bibnamefont {Pitaevskii}}, \ and\
  \bibinfo {author} {\bibfnamefont {S.}~\bibnamefont {Stringari}},\ }\href@noop
  {} {\bibfield  {journal} {\bibinfo  {journal} {Rev. Mod. Phys.}\ }\textbf
  {\bibinfo {volume} {71}},\ \bibinfo {pages} {463} (\bibinfo {year}
  {1999})}\BibitemShut {NoStop}%
\bibitem [{\citenamefont {Jin}\ \emph {et~al.}(1996)\citenamefont {Jin},
  \citenamefont {Ensher}, \citenamefont {Matthews}, \citenamefont {Wieman},\
  and\ \citenamefont {Cornell}}]{jin1996collective}%
  \BibitemOpen
  \bibfield  {author} {\bibinfo {author} {\bibfnamefont {D.~S.}\ \bibnamefont
  {Jin}}, \bibinfo {author} {\bibfnamefont {J.~R.}\ \bibnamefont {Ensher}},
  \bibinfo {author} {\bibfnamefont {M.~R.}\ \bibnamefont {Matthews}}, \bibinfo
  {author} {\bibfnamefont {C.~E.}\ \bibnamefont {Wieman}}, \ and\ \bibinfo
  {author} {\bibfnamefont {E.~A.}\ \bibnamefont {Cornell}},\ }\href@noop {}
  {\bibfield  {journal} {\bibinfo  {journal} {Phys. Rev. Lett.}\ }\textbf
  {\bibinfo {volume} {77}},\ \bibinfo {pages} {420} (\bibinfo {year}
  {1996})}\BibitemShut {NoStop}%
\bibitem [{\citenamefont {Pollack}\ \emph {et~al.}(2010)\citenamefont
  {Pollack}, \citenamefont {Dries}, \citenamefont {Hulet}, \citenamefont
  {Magalh\~aes}, \citenamefont {Henn}, \citenamefont {Ramos}, \citenamefont
  {Caracanhas},\ and\ \citenamefont {Bagnato}}]{Pollack2010collective}%
  \BibitemOpen
  \bibfield  {author} {\bibinfo {author} {\bibfnamefont {S.~E.}\ \bibnamefont
  {Pollack}}, \bibinfo {author} {\bibfnamefont {D.}~\bibnamefont {Dries}},
  \bibinfo {author} {\bibfnamefont {R.~G.}\ \bibnamefont {Hulet}}, \bibinfo
  {author} {\bibfnamefont {K.~M.~F.}\ \bibnamefont {Magalh\~aes}}, \bibinfo
  {author} {\bibfnamefont {E.~A.~L.}\ \bibnamefont {Henn}}, \bibinfo {author}
  {\bibfnamefont {E.~R.~F.}\ \bibnamefont {Ramos}}, \bibinfo {author}
  {\bibfnamefont {M.~A.}\ \bibnamefont {Caracanhas}}, \ and\ \bibinfo {author}
  {\bibfnamefont {V.~S.}\ \bibnamefont {Bagnato}},\ }\href@noop {} {\bibfield
  {journal} {\bibinfo  {journal} {Phys. Rev. A}\ }\textbf {\bibinfo {volume}
  {81}},\ \bibinfo {pages} {053627} (\bibinfo {year} {2010})}\BibitemShut
  {NoStop}%
\bibitem [{\citenamefont {Skov}\ \emph {et~al.}(2020)\citenamefont {Skov},
  \citenamefont {Skou}, \citenamefont {J\o{}rgensen},\ and\ \citenamefont
  {Arlt}}]{skov2020lhy}%
  \BibitemOpen
  \bibfield  {author} {\bibinfo {author} {\bibfnamefont {T.~G.}\ \bibnamefont
  {Skov}}, \bibinfo {author} {\bibfnamefont {M.~G.}\ \bibnamefont {Skou}},
  \bibinfo {author} {\bibfnamefont {N.~B.}\ \bibnamefont {J\o{}rgensen}}, \
  and\ \bibinfo {author} {\bibfnamefont {J.~J.}\ \bibnamefont {Arlt}},\
  }\href@noop {} {\bibfield  {journal} {\bibinfo  {journal} {arXiv:2011.02745}\
  } (\bibinfo {year} {2020})}\BibitemShut {NoStop}%
\bibitem [{\citenamefont {Atas}\ \emph {et~al.}(2017)\citenamefont {Atas},
  \citenamefont {Bouchoule}, \citenamefont {Gangardt},\ and\ \citenamefont
  {Kheruntsyan}}]{Atas2017}%
  \BibitemOpen
  \bibfield  {author} {\bibinfo {author} {\bibfnamefont {Y.~Y.}\ \bibnamefont
  {Atas}}, \bibinfo {author} {\bibfnamefont {I.}~\bibnamefont {Bouchoule}},
  \bibinfo {author} {\bibfnamefont {D.~M.}\ \bibnamefont {Gangardt}}, \ and\
  \bibinfo {author} {\bibfnamefont {K.~V.}\ \bibnamefont {Kheruntsyan}},\
  }\href@noop {} {\bibfield  {journal} {\bibinfo  {journal} {Phys. Rev. A}\
  }\textbf {\bibinfo {volume} {96}},\ \bibinfo {pages} {041605} (\bibinfo
  {year} {2017})}\BibitemShut {NoStop}%
\bibitem [{\citenamefont {De~Rosi}\ and\ \citenamefont
  {Stringari}(2015)}]{rosi2015collective}%
  \BibitemOpen
  \bibfield  {author} {\bibinfo {author} {\bibfnamefont {G.}~\bibnamefont
  {De~Rosi}}\ and\ \bibinfo {author} {\bibfnamefont {S.}~\bibnamefont
  {Stringari}},\ }\href@noop {} {\bibfield  {journal} {\bibinfo  {journal}
  {Physical Review A}\ }\textbf {\bibinfo {volume} {92}},\ \bibinfo {pages}
  {053617} (\bibinfo {year} {2015})}\BibitemShut {NoStop}%
\bibitem [{\citenamefont {Kinast}\ \emph {et~al.}(2004)\citenamefont {Kinast},
  \citenamefont {Hemmer}, \citenamefont {Gehm}, \citenamefont {Turlapov},\ and\
  \citenamefont {Thomas}}]{Kinast2004evidence}%
  \BibitemOpen
  \bibfield  {author} {\bibinfo {author} {\bibfnamefont {J.}~\bibnamefont
  {Kinast}}, \bibinfo {author} {\bibfnamefont {S.~L.}\ \bibnamefont {Hemmer}},
  \bibinfo {author} {\bibfnamefont {M.~E.}\ \bibnamefont {Gehm}}, \bibinfo
  {author} {\bibfnamefont {A.}~\bibnamefont {Turlapov}}, \ and\ \bibinfo
  {author} {\bibfnamefont {J.~E.}\ \bibnamefont {Thomas}},\ }\href@noop {}
  {\bibfield  {journal} {\bibinfo  {journal} {Phys. Rev. Lett.}\ }\textbf
  {\bibinfo {volume} {92}},\ \bibinfo {pages} {150402} (\bibinfo {year}
  {2004})}\BibitemShut {NoStop}%
\bibitem [{\citenamefont {Bartenstein}\ \emph {et~al.}(2004)\citenamefont
  {Bartenstein}, \citenamefont {Altmeyer}, \citenamefont {Riedl}, \citenamefont
  {Jochim}, \citenamefont {Chin}, \citenamefont {Denschlag},\ and\
  \citenamefont {Grimm}}]{Bartenstein2004collective}%
  \BibitemOpen
  \bibfield  {author} {\bibinfo {author} {\bibfnamefont {M.}~\bibnamefont
  {Bartenstein}}, \bibinfo {author} {\bibfnamefont {A.}~\bibnamefont
  {Altmeyer}}, \bibinfo {author} {\bibfnamefont {S.}~\bibnamefont {Riedl}},
  \bibinfo {author} {\bibfnamefont {S.}~\bibnamefont {Jochim}}, \bibinfo
  {author} {\bibfnamefont {C.}~\bibnamefont {Chin}}, \bibinfo {author}
  {\bibfnamefont {J.~H.}\ \bibnamefont {Denschlag}}, \ and\ \bibinfo {author}
  {\bibfnamefont {R.}~\bibnamefont {Grimm}},\ }\href@noop {} {\bibfield
  {journal} {\bibinfo  {journal} {Phys. Rev. Lett.}\ }\textbf {\bibinfo
  {volume} {92}},\ \bibinfo {pages} {203201} (\bibinfo {year}
  {2004})}\BibitemShut {NoStop}%
\bibitem [{\citenamefont {Altmeyer}\ \emph {et~al.}(2007)\citenamefont
  {Altmeyer}, \citenamefont {Riedl}, \citenamefont {Kohstall}, \citenamefont
  {Wright}, \citenamefont {Geursen}, \citenamefont {Bartenstein}, \citenamefont
  {Chin}, \citenamefont {Denschlag},\ and\ \citenamefont
  {Grimm}}]{altmeyer2007precision}%
  \BibitemOpen
  \bibfield  {author} {\bibinfo {author} {\bibfnamefont {A.}~\bibnamefont
  {Altmeyer}}, \bibinfo {author} {\bibfnamefont {S.}~\bibnamefont {Riedl}},
  \bibinfo {author} {\bibfnamefont {C.}~\bibnamefont {Kohstall}}, \bibinfo
  {author} {\bibfnamefont {M.~J.}\ \bibnamefont {Wright}}, \bibinfo {author}
  {\bibfnamefont {R.}~\bibnamefont {Geursen}}, \bibinfo {author} {\bibfnamefont
  {M.}~\bibnamefont {Bartenstein}}, \bibinfo {author} {\bibfnamefont
  {C.}~\bibnamefont {Chin}}, \bibinfo {author} {\bibfnamefont {J.~H.}\
  \bibnamefont {Denschlag}}, \ and\ \bibinfo {author} {\bibfnamefont
  {R.}~\bibnamefont {Grimm}},\ }\href@noop {} {\bibfield  {journal} {\bibinfo
  {journal} {Phys. Rev. Lett.}\ }\textbf {\bibinfo {volume} {98}},\ \bibinfo
  {pages} {040401} (\bibinfo {year} {2007})}\BibitemShut {NoStop}%
\bibitem [{\citenamefont {Tylutki}\ \emph {et~al.}(2020)\citenamefont
  {Tylutki}, \citenamefont {Astrakharchik}, \citenamefont {Malomed},\ and\
  \citenamefont {Petrov}}]{tylutki2020collective}%
  \BibitemOpen
  \bibfield  {author} {\bibinfo {author} {\bibfnamefont {M.}~\bibnamefont
  {Tylutki}}, \bibinfo {author} {\bibfnamefont {G.~E.}\ \bibnamefont
  {Astrakharchik}}, \bibinfo {author} {\bibfnamefont {B.~A.}\ \bibnamefont
  {Malomed}}, \ and\ \bibinfo {author} {\bibfnamefont {D.~S.}\ \bibnamefont
  {Petrov}},\ }\href@noop {} {\bibfield  {journal} {\bibinfo  {journal} {Phys.
  Rev. A}\ }\textbf {\bibinfo {volume} {101}},\ \bibinfo {pages} {051601}
  (\bibinfo {year} {2020})}\BibitemShut {NoStop}%
\bibitem [{\citenamefont {Hu}\ and\ \citenamefont
  {Liu}(2020{\natexlab{a}})}]{hu2020collective}%
  \BibitemOpen
  \bibfield  {author} {\bibinfo {author} {\bibfnamefont {H.}~\bibnamefont
  {Hu}}\ and\ \bibinfo {author} {\bibfnamefont {X.-J.}\ \bibnamefont {Liu}},\
  }\href@noop {} {\bibfield  {journal} {\bibinfo  {journal} {Physical Review
  A}\ }\textbf {\bibinfo {volume} {102}},\ \bibinfo {pages} {053303} (\bibinfo
  {year} {2020}{\natexlab{a}})}\BibitemShut {NoStop}%
\bibitem [{\citenamefont {Tanzi}\ \emph {et~al.}(2019)\citenamefont {Tanzi},
  \citenamefont {Roccuzzo}, \citenamefont {Lucioni}, \citenamefont {Fam{\`a}},
  \citenamefont {Fioretti}, \citenamefont {Gabbanini}, \citenamefont {Modugno},
  \citenamefont {Recati},\ and\ \citenamefont
  {Stringari}}]{tanzi2019supersolid}%
  \BibitemOpen
  \bibfield  {author} {\bibinfo {author} {\bibfnamefont {L.}~\bibnamefont
  {Tanzi}}, \bibinfo {author} {\bibfnamefont {S.}~\bibnamefont {Roccuzzo}},
  \bibinfo {author} {\bibfnamefont {E.}~\bibnamefont {Lucioni}}, \bibinfo
  {author} {\bibfnamefont {F.}~\bibnamefont {Fam{\`a}}}, \bibinfo {author}
  {\bibfnamefont {A.}~\bibnamefont {Fioretti}}, \bibinfo {author}
  {\bibfnamefont {C.}~\bibnamefont {Gabbanini}}, \bibinfo {author}
  {\bibfnamefont {G.}~\bibnamefont {Modugno}}, \bibinfo {author} {\bibfnamefont
  {A.}~\bibnamefont {Recati}}, \ and\ \bibinfo {author} {\bibfnamefont
  {S.}~\bibnamefont {Stringari}},\ }\href@noop {} {\bibfield  {journal}
  {\bibinfo  {journal} {Nature}\ }\textbf {\bibinfo {volume} {574}},\ \bibinfo
  {pages} {382} (\bibinfo {year} {2019})}\BibitemShut {NoStop}%
\bibitem [{\citenamefont {L{\'e}onard}\ \emph {et~al.}(2017)\citenamefont
  {L{\'e}onard}, \citenamefont {Morales}, \citenamefont {Zupancic},
  \citenamefont {Donner},\ and\ \citenamefont
  {Esslinger}}]{leonard2017monitoring}%
  \BibitemOpen
  \bibfield  {author} {\bibinfo {author} {\bibfnamefont {J.}~\bibnamefont
  {L{\'e}onard}}, \bibinfo {author} {\bibfnamefont {A.}~\bibnamefont
  {Morales}}, \bibinfo {author} {\bibfnamefont {P.}~\bibnamefont {Zupancic}},
  \bibinfo {author} {\bibfnamefont {T.}~\bibnamefont {Donner}}, \ and\ \bibinfo
  {author} {\bibfnamefont {T.}~\bibnamefont {Esslinger}},\ }\href@noop {}
  {\bibfield  {journal} {\bibinfo  {journal} {Science}\ }\textbf {\bibinfo
  {volume} {358}},\ \bibinfo {pages} {1415} (\bibinfo {year}
  {2017})}\BibitemShut {NoStop}%
\bibitem [{\citenamefont {Petrov}(2015)}]{petrov2015quantum}%
  \BibitemOpen
  \bibfield  {author} {\bibinfo {author} {\bibfnamefont {D.~S.}\ \bibnamefont
  {Petrov}},\ }\href@noop {} {\bibfield  {journal} {\bibinfo  {journal} {Phys.
  Rev. Lett.}\ }\textbf {\bibinfo {volume} {115}},\ \bibinfo {pages} {155302}
  (\bibinfo {year} {2015})}\BibitemShut {NoStop}%
\bibitem [{\citenamefont {Cabrera}\ \emph {et~al.}(2018)\citenamefont
  {Cabrera}, \citenamefont {Tanzi}, \citenamefont {Sanz}, \citenamefont
  {Naylor}, \citenamefont {Thomas}, \citenamefont {Cheiney},\ and\
  \citenamefont {Tarruell}}]{cabrera2018quantum}%
  \BibitemOpen
  \bibfield  {author} {\bibinfo {author} {\bibfnamefont {C.}~\bibnamefont
  {Cabrera}}, \bibinfo {author} {\bibfnamefont {L.}~\bibnamefont {Tanzi}},
  \bibinfo {author} {\bibfnamefont {J.}~\bibnamefont {Sanz}}, \bibinfo {author}
  {\bibfnamefont {B.}~\bibnamefont {Naylor}}, \bibinfo {author} {\bibfnamefont
  {P.}~\bibnamefont {Thomas}}, \bibinfo {author} {\bibfnamefont
  {P.}~\bibnamefont {Cheiney}}, \ and\ \bibinfo {author} {\bibfnamefont
  {L.}~\bibnamefont {Tarruell}},\ }\href@noop {} {\bibfield  {journal}
  {\bibinfo  {journal} {Science}\ }\textbf {\bibinfo {volume} {359}},\ \bibinfo
  {pages} {301} (\bibinfo {year} {2018})}\BibitemShut {NoStop}%
\bibitem [{\citenamefont {Cheiney}\ \emph {et~al.}(2018)\citenamefont
  {Cheiney}, \citenamefont {Cabrera}, \citenamefont {Sanz}, \citenamefont
  {Naylor}, \citenamefont {Tanzi},\ and\ \citenamefont
  {Tarruell}}]{cheiney2018bright}%
  \BibitemOpen
  \bibfield  {author} {\bibinfo {author} {\bibfnamefont {P.}~\bibnamefont
  {Cheiney}}, \bibinfo {author} {\bibfnamefont {C.~R.}\ \bibnamefont
  {Cabrera}}, \bibinfo {author} {\bibfnamefont {J.}~\bibnamefont {Sanz}},
  \bibinfo {author} {\bibfnamefont {B.}~\bibnamefont {Naylor}}, \bibinfo
  {author} {\bibfnamefont {L.}~\bibnamefont {Tanzi}}, \ and\ \bibinfo {author}
  {\bibfnamefont {L.}~\bibnamefont {Tarruell}},\ }\href@noop {} {\bibfield
  {journal} {\bibinfo  {journal} {Phys. Rev. Lett.}\ }\textbf {\bibinfo
  {volume} {120}},\ \bibinfo {pages} {135301} (\bibinfo {year}
  {2018})}\BibitemShut {NoStop}%
\bibitem [{\citenamefont {Semeghini}\ \emph {et~al.}(2018)\citenamefont
  {Semeghini}, \citenamefont {Ferioli}, \citenamefont {Masi}, \citenamefont
  {Mazzinghi}, \citenamefont {Wolswijk}, \citenamefont {Minardi}, \citenamefont
  {Modugno}, \citenamefont {Modugno}, \citenamefont {Inguscio},\ and\
  \citenamefont {Fattori}}]{semeghini2018self}%
  \BibitemOpen
  \bibfield  {author} {\bibinfo {author} {\bibfnamefont {G.}~\bibnamefont
  {Semeghini}}, \bibinfo {author} {\bibfnamefont {G.}~\bibnamefont {Ferioli}},
  \bibinfo {author} {\bibfnamefont {L.}~\bibnamefont {Masi}}, \bibinfo {author}
  {\bibfnamefont {C.}~\bibnamefont {Mazzinghi}}, \bibinfo {author}
  {\bibfnamefont {L.}~\bibnamefont {Wolswijk}}, \bibinfo {author}
  {\bibfnamefont {F.}~\bibnamefont {Minardi}}, \bibinfo {author} {\bibfnamefont
  {M.}~\bibnamefont {Modugno}}, \bibinfo {author} {\bibfnamefont
  {G.}~\bibnamefont {Modugno}}, \bibinfo {author} {\bibfnamefont
  {M.}~\bibnamefont {Inguscio}}, \ and\ \bibinfo {author} {\bibfnamefont
  {M.}~\bibnamefont {Fattori}},\ }\href@noop {} {\bibfield  {journal} {\bibinfo
   {journal} {Phys. Rev. Lett.}\ }\textbf {\bibinfo {volume} {120}},\ \bibinfo
  {pages} {235301} (\bibinfo {year} {2018})}\BibitemShut {NoStop}%
\bibitem [{\citenamefont {D'Errico}\ \emph {et~al.}(2019)\citenamefont
  {D'Errico}, \citenamefont {Burchianti}, \citenamefont {Prevedelli},
  \citenamefont {Salasnich}, \citenamefont {Ancilotto}, \citenamefont
  {Modugno}, \citenamefont {Minardi},\ and\ \citenamefont
  {Fort}}]{derrico2019observation}%
  \BibitemOpen
  \bibfield  {author} {\bibinfo {author} {\bibfnamefont {C.}~\bibnamefont
  {D'Errico}}, \bibinfo {author} {\bibfnamefont {A.}~\bibnamefont
  {Burchianti}}, \bibinfo {author} {\bibfnamefont {M.}~\bibnamefont
  {Prevedelli}}, \bibinfo {author} {\bibfnamefont {L.}~\bibnamefont
  {Salasnich}}, \bibinfo {author} {\bibfnamefont {F.}~\bibnamefont
  {Ancilotto}}, \bibinfo {author} {\bibfnamefont {M.}~\bibnamefont {Modugno}},
  \bibinfo {author} {\bibfnamefont {F.}~\bibnamefont {Minardi}}, \ and\
  \bibinfo {author} {\bibfnamefont {C.}~\bibnamefont {Fort}},\ }\href@noop {}
  {\bibfield  {journal} {\bibinfo  {journal} {Phys. Rev. Research}\ }\textbf
  {\bibinfo {volume} {1}},\ \bibinfo {pages} {033155} (\bibinfo {year}
  {2019})}\BibitemShut {NoStop}%
\bibitem [{\citenamefont {Lee}\ and\ \citenamefont {Yang}(1957)}]{lee1957many}%
  \BibitemOpen
  \bibfield  {author} {\bibinfo {author} {\bibfnamefont {T.~D.}\ \bibnamefont
  {Lee}}\ and\ \bibinfo {author} {\bibfnamefont {C.~N.}\ \bibnamefont {Yang}},\
  }\href@noop {} {\bibfield  {journal} {\bibinfo  {journal} {Phys. Rev.}\
  }\textbf {\bibinfo {volume} {105}},\ \bibinfo {pages} {1119} (\bibinfo {year}
  {1957})}\BibitemShut {NoStop}%
\bibitem [{\citenamefont {Huang}\ and\ \citenamefont
  {Yang}(1957)}]{huang1957quantum}%
  \BibitemOpen
  \bibfield  {author} {\bibinfo {author} {\bibfnamefont {K.}~\bibnamefont
  {Huang}}\ and\ \bibinfo {author} {\bibfnamefont {C.~N.}\ \bibnamefont
  {Yang}},\ }\href@noop {} {\bibfield  {journal} {\bibinfo  {journal} {Phys.
  Rev.}\ }\textbf {\bibinfo {volume} {105}},\ \bibinfo {pages} {767} (\bibinfo
  {year} {1957})}\BibitemShut {NoStop}%
\bibitem [{\citenamefont {Larsen}(1963)}]{larsen1963binary}%
  \BibitemOpen
  \bibfield  {author} {\bibinfo {author} {\bibfnamefont {D.~M.}\ \bibnamefont
  {Larsen}},\ }\href@noop {} {\bibfield  {journal} {\bibinfo  {journal} {Annals
  of Physics}\ }\textbf {\bibinfo {volume} {24}},\ \bibinfo {pages} {89 }
  (\bibinfo {year} {1963})}\BibitemShut {NoStop}%
\bibitem [{\citenamefont {Minardi}\ \emph {et~al.}(2019)\citenamefont
  {Minardi}, \citenamefont {Ancilotto}, \citenamefont {Burchianti},
  \citenamefont {D'Errico}, \citenamefont {Fort},\ and\ \citenamefont
  {Modugno}}]{minardi2019effective}%
  \BibitemOpen
  \bibfield  {author} {\bibinfo {author} {\bibfnamefont {F.}~\bibnamefont
  {Minardi}}, \bibinfo {author} {\bibfnamefont {F.}~\bibnamefont {Ancilotto}},
  \bibinfo {author} {\bibfnamefont {A.}~\bibnamefont {Burchianti}}, \bibinfo
  {author} {\bibfnamefont {C.}~\bibnamefont {D'Errico}}, \bibinfo {author}
  {\bibfnamefont {C.}~\bibnamefont {Fort}}, \ and\ \bibinfo {author}
  {\bibfnamefont {M.}~\bibnamefont {Modugno}},\ }\href@noop {} {\bibfield
  {journal} {\bibinfo  {journal} {Phys. Rev. A}\ }\textbf {\bibinfo {volume}
  {100}},\ \bibinfo {pages} {063636} (\bibinfo {year} {2019})}\BibitemShut
  {NoStop}%
\bibitem [{\citenamefont {Naidon}\ and\ \citenamefont
  {Petrov}(2021)}]{naidon2021bubbles}%
  \BibitemOpen
  \bibfield  {author} {\bibinfo {author} {\bibfnamefont {P.}~\bibnamefont
  {Naidon}}\ and\ \bibinfo {author} {\bibfnamefont {D.~S.}\ \bibnamefont
  {Petrov}},\ }\href@noop {} {\bibfield  {journal} {\bibinfo  {journal} {Phys.
  Rev. Lett.}\ }\textbf {\bibinfo {volume} {126}},\ \bibinfo {pages} {115301}
  (\bibinfo {year} {2021})}\BibitemShut {NoStop}%
\bibitem [{\citenamefont {Petrov}\ and\ \citenamefont
  {Astrakharchik}(2016)}]{petrov2016ultradilute}%
  \BibitemOpen
  \bibfield  {author} {\bibinfo {author} {\bibfnamefont {D.~S.}\ \bibnamefont
  {Petrov}}\ and\ \bibinfo {author} {\bibfnamefont {G.~E.}\ \bibnamefont
  {Astrakharchik}},\ }\href@noop {} {\bibfield  {journal} {\bibinfo  {journal}
  {Phys. Rev. Lett.}\ }\textbf {\bibinfo {volume} {117}},\ \bibinfo {pages}
  {100401} (\bibinfo {year} {2016})}\BibitemShut {NoStop}%
\bibitem [{\citenamefont {Parisi}\ \emph {et~al.}(2019)\citenamefont {Parisi},
  \citenamefont {Astrakharchik},\ and\ \citenamefont
  {Giorgini}}]{parisi2019liquid}%
  \BibitemOpen
  \bibfield  {author} {\bibinfo {author} {\bibfnamefont {L.}~\bibnamefont
  {Parisi}}, \bibinfo {author} {\bibfnamefont {G.~E.}\ \bibnamefont
  {Astrakharchik}}, \ and\ \bibinfo {author} {\bibfnamefont {S.}~\bibnamefont
  {Giorgini}},\ }\href@noop {} {\bibfield  {journal} {\bibinfo  {journal}
  {Phys. Rev. Lett.}\ }\textbf {\bibinfo {volume} {122}},\ \bibinfo {pages}
  {105302} (\bibinfo {year} {2019})}\BibitemShut {NoStop}%
\bibitem [{\citenamefont {Parisi}\ and\ \citenamefont
  {Giorgini}(2020)}]{parisi2020quantum}%
  \BibitemOpen
  \bibfield  {author} {\bibinfo {author} {\bibfnamefont {L.}~\bibnamefont
  {Parisi}}\ and\ \bibinfo {author} {\bibfnamefont {S.}~\bibnamefont
  {Giorgini}},\ }\href@noop {} {\bibfield  {journal} {\bibinfo  {journal}
  {Phys. Rev. A}\ }\textbf {\bibinfo {volume} {102}},\ \bibinfo {pages}
  {023318} (\bibinfo {year} {2020})}\BibitemShut {NoStop}%
\bibitem [{\citenamefont {B\"ottcher}\ \emph {et~al.}(2019)\citenamefont
  {B\"ottcher}, \citenamefont {Wenzel}, \citenamefont {Schmidt}, \citenamefont
  {Guo}, \citenamefont {Langen}, \citenamefont {Ferrier-Barbut}, \citenamefont
  {Pfau}, \citenamefont {Bomb\'{\i}n}, \citenamefont {S\'anchez-Baena},
  \citenamefont {Boronat},\ and\ \citenamefont
  {Mazzanti}}]{bottcher2019dilute}%
  \BibitemOpen
  \bibfield  {author} {\bibinfo {author} {\bibfnamefont {F.}~\bibnamefont
  {B\"ottcher}}, \bibinfo {author} {\bibfnamefont {M.}~\bibnamefont {Wenzel}},
  \bibinfo {author} {\bibfnamefont {J.-N.}\ \bibnamefont {Schmidt}}, \bibinfo
  {author} {\bibfnamefont {M.}~\bibnamefont {Guo}}, \bibinfo {author}
  {\bibfnamefont {T.}~\bibnamefont {Langen}}, \bibinfo {author} {\bibfnamefont
  {I.}~\bibnamefont {Ferrier-Barbut}}, \bibinfo {author} {\bibfnamefont
  {T.}~\bibnamefont {Pfau}}, \bibinfo {author} {\bibfnamefont {R.}~\bibnamefont
  {Bomb\'{\i}n}}, \bibinfo {author} {\bibfnamefont {J.}~\bibnamefont
  {S\'anchez-Baena}}, \bibinfo {author} {\bibfnamefont {J.}~\bibnamefont
  {Boronat}}, \ and\ \bibinfo {author} {\bibfnamefont {F.}~\bibnamefont
  {Mazzanti}},\ }\href@noop {} {\bibfield  {journal} {\bibinfo  {journal}
  {Phys. Rev. Research}\ }\textbf {\bibinfo {volume} {1}},\ \bibinfo {pages}
  {033088} (\bibinfo {year} {2019})}\BibitemShut {NoStop}%
\bibitem [{\citenamefont {Böttcher}\ \emph {et~al.}(2021)\citenamefont
  {Böttcher}, \citenamefont {Schmidt}, \citenamefont {Hertkorn}, \citenamefont
  {Ng}, \citenamefont {Graham}, \citenamefont {Guo}, \citenamefont {Langen},\
  and\ \citenamefont {Pfau}}]{bottcher2021}%
  \BibitemOpen
  \bibfield  {author} {\bibinfo {author} {\bibfnamefont {F.}~\bibnamefont
  {Böttcher}}, \bibinfo {author} {\bibfnamefont {J.-N.}\ \bibnamefont
  {Schmidt}}, \bibinfo {author} {\bibfnamefont {J.}~\bibnamefont {Hertkorn}},
  \bibinfo {author} {\bibfnamefont {K.~S.~H.}\ \bibnamefont {Ng}}, \bibinfo
  {author} {\bibfnamefont {S.~D.}\ \bibnamefont {Graham}}, \bibinfo {author}
  {\bibfnamefont {M.}~\bibnamefont {Guo}}, \bibinfo {author} {\bibfnamefont
  {T.}~\bibnamefont {Langen}}, \ and\ \bibinfo {author} {\bibfnamefont
  {T.}~\bibnamefont {Pfau}},\ }\href@noop {} {\bibfield  {journal} {\bibinfo
  {journal} {Reports on Progress in Physics}\ }\textbf {\bibinfo {volume}
  {84}},\ \bibinfo {pages} {012403} (\bibinfo {year} {2021})}\BibitemShut
  {NoStop}%
\bibitem [{\citenamefont {Barranco}\ \emph {et~al.}(2006)\citenamefont
  {Barranco}, \citenamefont {Guardiola}, \citenamefont {Hern\'andez},
  \citenamefont {Mayol}, \citenamefont {Navarro},\ and\ \citenamefont
  {Pi}}]{barranco2006helium}%
  \BibitemOpen
  \bibfield  {author} {\bibinfo {author} {\bibfnamefont {M.}~\bibnamefont
  {Barranco}}, \bibinfo {author} {\bibfnamefont {R.}~\bibnamefont {Guardiola}},
  \bibinfo {author} {\bibfnamefont {S.}~\bibnamefont {Hern\'andez}}, \bibinfo
  {author} {\bibfnamefont {R.}~\bibnamefont {Mayol}}, \bibinfo {author}
  {\bibfnamefont {J.}~\bibnamefont {Navarro}}, \ and\ \bibinfo {author}
  {\bibfnamefont {M.}~\bibnamefont {Pi}},\ }\href@noop {} {\bibfield  {journal}
  {\bibinfo  {journal} {J. Low Temp. Phys}\ }\textbf {\bibinfo {volume}
  {142}},\ \bibinfo {pages} {1} (\bibinfo {year} {2006})}\BibitemShut {NoStop}%
\bibitem [{\citenamefont {Hu}\ and\ \citenamefont
  {Liu}(2020{\natexlab{b}})}]{hu2020consistent}%
  \BibitemOpen
  \bibfield  {author} {\bibinfo {author} {\bibfnamefont {H.}~\bibnamefont
  {Hu}}\ and\ \bibinfo {author} {\bibfnamefont {X.-J.}\ \bibnamefont {Liu}},\
  }\href@noop {} {\bibfield  {journal} {\bibinfo  {journal} {Phys. Rev. Lett.}\
  }\textbf {\bibinfo {volume} {125}},\ \bibinfo {pages} {195302} (\bibinfo
  {year} {2020}{\natexlab{b}})}\BibitemShut {NoStop}%
\bibitem [{\citenamefont {Giorgini}\ \emph {et~al.}(1999)\citenamefont
  {Giorgini}, \citenamefont {Boronat},\ and\ \citenamefont
  {Casulleras}}]{giorgini1999ground}%
  \BibitemOpen
  \bibfield  {author} {\bibinfo {author} {\bibfnamefont {S.}~\bibnamefont
  {Giorgini}}, \bibinfo {author} {\bibfnamefont {J.}~\bibnamefont {Boronat}}, \
  and\ \bibinfo {author} {\bibfnamefont {J.}~\bibnamefont {Casulleras}},\
  }\href@noop {} {\bibfield  {journal} {\bibinfo  {journal} {Phys. Rev. A}\
  }\textbf {\bibinfo {volume} {60}},\ \bibinfo {pages} {5129} (\bibinfo {year}
  {1999})}\BibitemShut {NoStop}%
\bibitem [{\citenamefont {Rossi}\ and\ \citenamefont
  {Salasnich}(2013)}]{rossi2013path}%
  \BibitemOpen
  \bibfield  {author} {\bibinfo {author} {\bibfnamefont {M.}~\bibnamefont
  {Rossi}}\ and\ \bibinfo {author} {\bibfnamefont {L.}~\bibnamefont
  {Salasnich}},\ }\href@noop {} {\bibfield  {journal} {\bibinfo  {journal}
  {Phys. Rev. A}\ }\textbf {\bibinfo {volume} {88}},\ \bibinfo {pages} {053617}
  (\bibinfo {year} {2013})}\BibitemShut {NoStop}%
\bibitem [{\citenamefont {Cikojevi\'c}\ \emph {et~al.}(2020)\citenamefont
  {Cikojevi\'c}, \citenamefont {Marki\'c},\ and\ \citenamefont
  {Boronat}}]{cikojevic2020finite}%
  \BibitemOpen
  \bibfield  {author} {\bibinfo {author} {\bibfnamefont {V.}~\bibnamefont
  {Cikojevi\'c}}, \bibinfo {author} {\bibfnamefont {L.~V.}\ \bibnamefont
  {Marki\'c}}, \ and\ \bibinfo {author} {\bibfnamefont {J.}~\bibnamefont
  {Boronat}},\ }\href@noop {} {\bibfield  {journal} {\bibinfo  {journal} {New
  Journal of Physics}\ }\textbf {\bibinfo {volume} {22}},\ \bibinfo {pages}
  {053045} (\bibinfo {year} {2020})}\BibitemShut {NoStop}%
\bibitem [{\citenamefont {Cikojevi\'c}\ \emph {et~al.}(2019)\citenamefont
  {Cikojevi\'c}, \citenamefont {Marki\'c}, \citenamefont {Astrakharchik},\ and\
  \citenamefont {Boronat}}]{cikojevic2019universality}%
  \BibitemOpen
  \bibfield  {author} {\bibinfo {author} {\bibfnamefont {V.}~\bibnamefont
  {Cikojevi\'c}}, \bibinfo {author} {\bibfnamefont {L.~V.}\ \bibnamefont
  {Marki\'c}}, \bibinfo {author} {\bibfnamefont {G.~E.}\ \bibnamefont
  {Astrakharchik}}, \ and\ \bibinfo {author} {\bibfnamefont {J.}~\bibnamefont
  {Boronat}},\ }\href@noop {} {\bibfield  {journal} {\bibinfo  {journal} {Phys.
  Rev. A}\ }\textbf {\bibinfo {volume} {99}},\ \bibinfo {pages} {023618}
  (\bibinfo {year} {2019})}\BibitemShut {NoStop}%
\bibitem [{\citenamefont {Ferioli}\ \emph {et~al.}(2019)\citenamefont
  {Ferioli}, \citenamefont {Semeghini}, \citenamefont {Masi}, \citenamefont
  {Giusti}, \citenamefont {Modugno}, \citenamefont {Inguscio}, \citenamefont
  {Gallem\'{\i}}, \citenamefont {Recati},\ and\ \citenamefont
  {Fattori}}]{ferioli2019collisions}%
  \BibitemOpen
  \bibfield  {author} {\bibinfo {author} {\bibfnamefont {G.}~\bibnamefont
  {Ferioli}}, \bibinfo {author} {\bibfnamefont {G.}~\bibnamefont {Semeghini}},
  \bibinfo {author} {\bibfnamefont {L.}~\bibnamefont {Masi}}, \bibinfo {author}
  {\bibfnamefont {G.}~\bibnamefont {Giusti}}, \bibinfo {author} {\bibfnamefont
  {G.}~\bibnamefont {Modugno}}, \bibinfo {author} {\bibfnamefont
  {M.}~\bibnamefont {Inguscio}}, \bibinfo {author} {\bibfnamefont
  {A.}~\bibnamefont {Gallem\'{\i}}}, \bibinfo {author} {\bibfnamefont
  {A.}~\bibnamefont {Recati}}, \ and\ \bibinfo {author} {\bibfnamefont
  {M.}~\bibnamefont {Fattori}},\ }\href@noop {} {\bibfield  {journal} {\bibinfo
   {journal} {Phys. Rev. Lett.}\ }\textbf {\bibinfo {volume} {122}},\ \bibinfo
  {pages} {090401} (\bibinfo {year} {2019})}\BibitemShut {NoStop}%
\bibitem [{\citenamefont {Frohn}\ and\ \citenamefont
  {Roth}(2000)}]{frohn2000dynamics}%
  \BibitemOpen
  \bibfield  {author} {\bibinfo {author} {\bibfnamefont {A.}~\bibnamefont
  {Frohn}}\ and\ \bibinfo {author} {\bibfnamefont {N.}~\bibnamefont {Roth}},\
  }\href@noop {} {\emph {\bibinfo {title} {Dynamics of droplets}}}\ (\bibinfo
  {publisher} {Springer Science \& Business Media},\ \bibinfo {year}
  {2000})\BibitemShut {NoStop}%
\bibitem [{\citenamefont {Staudinger}\ \emph {et~al.}(2018)\citenamefont
  {Staudinger}, \citenamefont {Mazzanti},\ and\ \citenamefont
  {Zillich}}]{staudinger2018self}%
  \BibitemOpen
  \bibfield  {author} {\bibinfo {author} {\bibfnamefont {C.}~\bibnamefont
  {Staudinger}}, \bibinfo {author} {\bibfnamefont {F.}~\bibnamefont
  {Mazzanti}}, \ and\ \bibinfo {author} {\bibfnamefont {R.~E.}\ \bibnamefont
  {Zillich}},\ }\href@noop {} {\bibfield  {journal} {\bibinfo  {journal} {Phys.
  Rev. A}\ }\textbf {\bibinfo {volume} {98}},\ \bibinfo {pages} {023633}
  (\bibinfo {year} {2018})}\BibitemShut {NoStop}%
\bibitem [{\citenamefont {Ancilotto}\ \emph {et~al.}(2018)\citenamefont
  {Ancilotto}, \citenamefont {Barranco}, \citenamefont {Guilleumas},\ and\
  \citenamefont {Pi}}]{ancilotto2018self}%
  \BibitemOpen
  \bibfield  {author} {\bibinfo {author} {\bibfnamefont {F.}~\bibnamefont
  {Ancilotto}}, \bibinfo {author} {\bibfnamefont {M.}~\bibnamefont {Barranco}},
  \bibinfo {author} {\bibfnamefont {M.}~\bibnamefont {Guilleumas}}, \ and\
  \bibinfo {author} {\bibfnamefont {M.}~\bibnamefont {Pi}},\ }\href@noop {}
  {\bibfield  {journal} {\bibinfo  {journal} {Phys. Rev. A}\ }\textbf {\bibinfo
  {volume} {98}},\ \bibinfo {pages} {053623} (\bibinfo {year}
  {2018})}\BibitemShut {NoStop}%
\bibitem [{\citenamefont {Tanzi}\ \emph {et~al.}(2018)\citenamefont {Tanzi},
  \citenamefont {Cabrera}, \citenamefont {Sanz}, \citenamefont {Cheiney},
  \citenamefont {Tomza},\ and\ \citenamefont {Tarruell}}]{tanzi2018feshbach}%
  \BibitemOpen
  \bibfield  {author} {\bibinfo {author} {\bibfnamefont {L.}~\bibnamefont
  {Tanzi}}, \bibinfo {author} {\bibfnamefont {C.~R.}\ \bibnamefont {Cabrera}},
  \bibinfo {author} {\bibfnamefont {J.}~\bibnamefont {Sanz}}, \bibinfo {author}
  {\bibfnamefont {P.}~\bibnamefont {Cheiney}}, \bibinfo {author} {\bibfnamefont
  {M.}~\bibnamefont {Tomza}}, \ and\ \bibinfo {author} {\bibfnamefont
  {L.}~\bibnamefont {Tarruell}},\ }\href@noop {} {\bibfield  {journal}
  {\bibinfo  {journal} {Phys. Rev. A}\ }\textbf {\bibinfo {volume} {98}},\
  \bibinfo {pages} {062712} (\bibinfo {year} {2018})}\BibitemShut {NoStop}%
\bibitem [{\citenamefont {Barranco}\ \emph {et~al.}(2017)\citenamefont
  {Barranco}, \citenamefont {Coppens}, \citenamefont {Halberstadt},
  \citenamefont {Hernando}, \citenamefont {Leal}, \citenamefont {Mateo},
  \citenamefont {Mayol},\ and\ \citenamefont {Pi}}]{barranco2017zero}%
  \BibitemOpen
  \bibfield  {author} {\bibinfo {author} {\bibfnamefont {M.}~\bibnamefont
  {Barranco}}, \bibinfo {author} {\bibfnamefont {F.}~\bibnamefont {Coppens}},
  \bibinfo {author} {\bibfnamefont {N.}~\bibnamefont {Halberstadt}}, \bibinfo
  {author} {\bibfnamefont {A.}~\bibnamefont {Hernando}}, \bibinfo {author}
  {\bibfnamefont {A.}~\bibnamefont {Leal}}, \bibinfo {author} {\bibfnamefont
  {D.}~\bibnamefont {Mateo}}, \bibinfo {author} {\bibfnamefont
  {R.}~\bibnamefont {Mayol}}, \ and\ \bibinfo {author} {\bibfnamefont
  {M.}~\bibnamefont {Pi}},\ }\href@noop {} {\enquote {\bibinfo {title} {{Zero
  temperature DFT and TDDFT for 4 He: A short guide for practitioners}},}\
  }\bibinfo {howpublished}
  {https://github.com/bcntls2016/DFT-Guide/blob/master/dft-guide.pdf} (\bibinfo
  {year} {2017})\BibitemShut {NoStop}%
\bibitem [{\citenamefont {Chin}\ \emph {et~al.}(2009)\citenamefont {Chin},
  \citenamefont {Janecek},\ and\ \citenamefont {Krotscheck}}]{chin2009any}%
  \BibitemOpen
  \bibfield  {author} {\bibinfo {author} {\bibfnamefont {S.~A.}\ \bibnamefont
  {Chin}}, \bibinfo {author} {\bibfnamefont {S.}~\bibnamefont {Janecek}}, \
  and\ \bibinfo {author} {\bibfnamefont {E.}~\bibnamefont {Krotscheck}},\
  }\href@noop {} {\bibfield  {journal} {\bibinfo  {journal} {Chemical Physics
  Letters}\ }\textbf {\bibinfo {volume} {470}},\ \bibinfo {pages} {342}
  (\bibinfo {year} {2009})}\BibitemShut {NoStop}%
\bibitem [{\citenamefont {Ferioli}\ \emph {et~al.}(2020)\citenamefont
  {Ferioli}, \citenamefont {Semeghini}, \citenamefont {Terradas-Brians\'o},
  \citenamefont {Masi}, \citenamefont {Fattori},\ and\ \citenamefont
  {Modugno}}]{ferioli2020dynamical}%
  \BibitemOpen
  \bibfield  {author} {\bibinfo {author} {\bibfnamefont {G.}~\bibnamefont
  {Ferioli}}, \bibinfo {author} {\bibfnamefont {G.}~\bibnamefont {Semeghini}},
  \bibinfo {author} {\bibfnamefont {S.}~\bibnamefont {Terradas-Brians\'o}},
  \bibinfo {author} {\bibfnamefont {L.}~\bibnamefont {Masi}}, \bibinfo {author}
  {\bibfnamefont {M.}~\bibnamefont {Fattori}}, \ and\ \bibinfo {author}
  {\bibfnamefont {M.}~\bibnamefont {Modugno}},\ }\href@noop {} {\bibfield
  {journal} {\bibinfo  {journal} {Phys. Rev. Research}\ }\textbf {\bibinfo
  {volume} {2}},\ \bibinfo {pages} {013269} (\bibinfo {year}
  {2020})}\BibitemShut {NoStop}%
\bibitem [{\citenamefont {Roy}\ \emph {et~al.}(2013)\citenamefont {Roy},
  \citenamefont {Landini}, \citenamefont {Trenkwalder}, \citenamefont
  {Semeghini}, \citenamefont {Spagnolli}, \citenamefont {Simoni}, \citenamefont
  {Fattori}, \citenamefont {Inguscio},\ and\ \citenamefont
  {Modugno}}]{roy2013test}%
  \BibitemOpen
  \bibfield  {author} {\bibinfo {author} {\bibfnamefont {S.}~\bibnamefont
  {Roy}}, \bibinfo {author} {\bibfnamefont {M.}~\bibnamefont {Landini}},
  \bibinfo {author} {\bibfnamefont {A.}~\bibnamefont {Trenkwalder}}, \bibinfo
  {author} {\bibfnamefont {G.}~\bibnamefont {Semeghini}}, \bibinfo {author}
  {\bibfnamefont {G.}~\bibnamefont {Spagnolli}}, \bibinfo {author}
  {\bibfnamefont {A.}~\bibnamefont {Simoni}}, \bibinfo {author} {\bibfnamefont
  {M.}~\bibnamefont {Fattori}}, \bibinfo {author} {\bibfnamefont
  {M.}~\bibnamefont {Inguscio}}, \ and\ \bibinfo {author} {\bibfnamefont
  {G.}~\bibnamefont {Modugno}},\ }\href@noop {} {\bibfield  {journal} {\bibinfo
   {journal} {Phys. Rev. Lett}\ }\textbf {\bibinfo {volume} {111}},\ \bibinfo
  {pages} {053202} (\bibinfo {year} {2013})}\BibitemShut {NoStop}%
\bibitem [{\citenamefont {material: Videos of the time-evolution of~selected
  collisions}()}]{supplement}%
  \BibitemOpen
  \bibfield  {author} {\bibinfo {author} {\bibfnamefont {S.}~\bibnamefont
  {material: Videos of the time-evolution of~selected collisions}},\
  }\href@noop {} {\ }\BibitemShut {NoStop}%
\bibitem [{\citenamefont {Wang}\ \emph {et~al.}(2020)\citenamefont {Wang},
  \citenamefont {Hu},\ and\ \citenamefont {Liu}}]{wang2020thermal}%
  \BibitemOpen
  \bibfield  {author} {\bibinfo {author} {\bibfnamefont {J.}~\bibnamefont
  {Wang}}, \bibinfo {author} {\bibfnamefont {H.}~\bibnamefont {Hu}}, \ and\
  \bibinfo {author} {\bibfnamefont {X.-J.}\ \bibnamefont {Liu}},\ }\href@noop
  {} {\bibfield  {journal} {\bibinfo  {journal} {New Journal of Physics}\
  }\textbf {\bibinfo {volume} {22}},\ \bibinfo {pages} {103044} (\bibinfo
  {year} {2020})}\BibitemShut {NoStop}%
\end{thebibliography}%
	
\end{document}